\documentclass[12pt]{article} 
\usepackage{amsmath}
\usepackage{slashed}
\usepackage[dvips]{graphicx}

\newcommand{\nn}{\nonumber}
\newcommand{\be}{\begin{equation}}
\newcommand{\ee}{\end{equation}}
\newcommand{\ba}{\begin{eqnarray}}
\newcommand{\ea}{\end{eqnarray}}
\newcommand{\ci}[1]{\cite{#1}}
\def\={\,=\,}

\def\als{\alpha_s}
\def\ale{\alpha_{\rm em}}
\def\mev{\,{\rm MeV}}
\def\gev{\,{\rm GeV}}

\newcommand{\sla}{\hspace*{-0.20cm}/}

\newcommand{\da}{{DA}}

\def\muR{\mu^2_R}     
\def\muF{\mu^2_F}
\def\muO{\mu^2_0}
\newcommand{\tw}{\textwidth}                          
\newcommand{\req}[1]{(\ref{#1})}

\def\eps{\epsilon}
\def\veps{\varepsilon}
\def\sh{\hat{s}}
\def\uh{\hat{u}}
\def\th{\hat{t}}

\def\taub{\bar{\tau}}

\def\phiDA{\phi}
\usepackage{amssymb}
\newcounter{comment}

{\refstepcounter{comment}%
\begin{quote}
\ttfamily\small$\blacksquare$ \textbf{\underline{Comment} $\sharp$\thecomment:}}%
{\end{quote}}

{%\refstepcounter{comment}
\begin{quote}
\ttfamily\small$\blacktriangleright$ \textbf{\underline{Reply} $\sharp$\thecomment:}}%
{\end{quote}}
\begin{document}
\thispagestyle{empty}
\begin{flushright}
%WU-B 18-00\\
  RBI-ThPhys-2020-54\\
  WUB/21-01\\
July, 09 2021\\[20mm]
\end{flushright}

\begin{center}

{\Large\bf Wide-angle photo- and electroproduction \\[0.3em]
of pions to twist-3 accuracy} \\
\vskip 15mm
P.\ Kroll \\[1em]
{\small {\it Fachbereich Physik, Universit\"at Wuppertal, D-42097 Wuppertal,
Germany}}\\[1em]
K.\  Passek-Kumeri\v{c}ki\\[1em]
{\small {\it Division of Theoretical Physics, Rudjer Bo\v{s}kovi\'{c} Institute, 
HR-10002 Zagreb, Croatia}}\\

\end{center}

\begin{abstract}
  We investigate wide-angle photo- and electroproduction of pions within the handbag mechanism in which
  the $\gamma^{(*)} N\to \pi N'$ amplitudes factorize into hard partonic subprocess amplitudes,
  $\gamma^{(*)} q\to \pi q'$,  and form factors representing $1/x$-moments of generalized parton
  distributions (GPDs). The subprocess is calculated to twist-3 accuracy taking into account
  the 2- and 3-body Fock components of the pion. Both components are required for achieving gauge
  invariance in QED and QCD. The twist-2 and twist-3 distribution amplitudes (DAs) of the pion as well
  as the form factors are taken from our study of $\pi^0$ photoproduction. Extensive results on
  photoproduction of charged pions are presented and compared to experiment. Predictions on
  electroproduction of pions as well as on spin effects are also given. As a byproduct of our analysis
  we also obtain the complete twist-3 subprocess amplitudes contributing to deeply-virtual
  electroproduction of pions.
\end{abstract}

\clearpage

\tableofcontents

\clearpage
%%%%%%%%%%%%%%%%%%%%%%%%%%%%%%%%%%%%%%%%%%%%%%%%%%%%%%%%%%%%%%%%%%%%%%
\section{Introduction}
\label{sec:introduction}
%%%%%%%%%%%%%%%%%%%%%%%%%%%%%%%%%%%%%%%%%%%%%%%%%%%%%%%%%%%%%%%%%%%%%%

Hard exclusive processes have attracted much attention in recent years. The main interest has been
focussed on the deeply virtual processes for which the virtuality of the photon provides
the hard scale. Examples for such processes are deep-virtual Compton scattering or meson
production (DVMP). The basis of the theoretical treatment of these processes is the representation
of the amplitudes as a convolution of a hard partonic subprocess and soft hadronic matrix
elements, parameterized as GPDs \ci{ji96,collins}. Besides the deeply virtual processes
also wide-angle ones have been investigated for which the hard scale is provided by a large
momentum transfer from the photon to the meson. Large momentum transfer is equivalent to large
Mandelstam variables $s, -t$ and $-u$.
The first process of this type, studied within this handbag mechanism, has been wide-angle Compton
scattering (WACS). There are reasonable arguments \ci{rad98,DFJK1} that, for this kinematical situation,
the Compton amplitude factorize into a product of perturbatively calculable partonic subprocess amplitudes
and form factors of the nucleon representing $1/x$-moments of zero-skewness GPDs. 
%Since the relevant GPDs in this case, $H$, $E$ and $\widetilde{H}$ for valence quarks are known 
%from an analysis of the nucleon's electromagnetic form factors \ci{DFJK4,DK13} one can compute 
%the Compton form factors and subsequently the Compton cross section. 
%The results of this parameter-free prediction \ci{DK13} agree fairly well with
%experiment \ci{danagoulian08}. 
Since the relevant GPDs in this case, $H$, $E$ and $\widetilde H$ at zero-skewness for
valence quarks are known from an analysis of the nucleon's electromagnetic form factors \ci{DFJK4,DK13} 
one can compute the Compton form factors and subsequently the Compton cross section. 
The results agree fairly well with experiment \ci{danagoulian08}. 
We stress that in this calculation no free parameter is fitted to the WACS data.
The authors of Ref. \ci{huang00} extended the approach advocated for in \ci{DFJK1}
to photoproduction of uncharged pions and $\rho$ mesons. 
It however turned out that the calculated
cross sections are way below the experimental data. An attempt \ci{signatures} to
improve this (twist-2) analysis by taking into account twist-3 contributions in the Wandzura-Wilczek
(WW) approximation, 
i.e.\ with only the $q\bar{q}$ Fock component of the pion considered, failed too - the
twist-3 contribution in WW approximation is zero in wide-angle photoproduction of pions. This is in
sharp contrast to DVMP where the WW approximation is non-zero \ci{GK5,GK6} and plays an important
role in the interpretation of the data on deeply virtual pion electroproduction \ci{hermes,clas12,defurne}.
With the arrival of the CLAS data  \ci{clas-pi0} on wide-angle photoproduction of $\pi^0$ mesons the
handbag mechanism has again been taken up by us \ci{KPK18}. Now, as a further development of the previous
study \ci{signatures}, the full twist-3 contribution to the subprocess amplitude is included in the analysis. 
Both its parts, the one from the $q\bar{q}$ Fock component of the pion as well as that from its
$q\bar{q}g$ Fock component, are connected by the equations of motion and both are required to accomplish
gauge invariance. The twist-3 contributions to the hard subprocess go along with leading-twist
transversity GPDs, $H_T$, $\bar{E}_T$ and $\widetilde{H}_T$ and the corresponding form factors. We stress
that this twist-3 effect, although being formally suppressed by $\mu_\pi/\sqrt{s}$ at the amplitude level,
is very strong due to the large mass parameter, $\mu_\pi$. As shown
in \ci{KPK18} the twist-3 contribution dominates the twist-2 one analogously to
the case of deeply virtual $\pi^0$ electroproduction \ci{GK6,clas12,defurne}. The outcome of \ci{KPK18} is
that the CLAS photoproduction data \ci{clas-pi0} can be fitted with reasonable pion \da s and soft form
factors. Twist-3 effects may also be generated by twist-3 GPDs \ci{belitsky} for which no 
enhancement similar to $\mu_\pi$ parameter is known. 
These twist-3 effects are expected to be small and are therefore neglected in \ci{KPK18}
as well as in the present work.

The present work is an extension of our previous analysis of wide-angle $\pi^0$ photoproduction \ci{KPK18}.
We now apply the handbag mechanism to twist-3 accuracy to wide-angle photoproduction of charged pions and,
in addition, we also investigate wide-angle electroproduction of pions. The latter process with its large
variety of observables may allow for a detailed study of the dynamics of this process and to fix the pion
\da s and form factors in detail. The large $-t$ behavior of the transversity form factors encodes the
large-t behavior of the transversity GPDs. The latter is required for the understanding of the parton
densities in the transverse position plane. We think that the present study is timely because of planned
experiments at the Jefferson laboratory.

The paper is organized as follows: In Sect.\ 2 we 
recapitulate  the handbag approach to photo- and electroproduction of pions to twist-3 accuracy. In 
the next section, Sect.\ 3, we present the results for the subprocess amplitudes and discuss
their properties as well as their photoproduction limit, $Q^2\to 0$, and the 
DVMP limit, $\th\to 0$, for which the subprocess amplitudes also hold. The soft
physics input to the handbag mechanism, namely the pion \da s and the form factors, are
discussed in Sect.\ 4. In the next section, Sect.\ 5, our results for photoproduction
of charged pions are presented and compared to experiment. In Sect.\ 6 we show predictions
for the partial cross sections of pion electroproduction. Sect.\ 7 is devoted to a discussion
of spin effects and in Sect.\ 8 remarks are made concerning the uncertainties of our predictions.
The paper ends with the usual summary. In an appendix we compile the familiar evolution properties
of the \da s.

%%%%%%%%%%%%%%%%%%%%%%%%%%%%%%%%%%%%%%%%%%%%%%%%%%%%%%%%%%%%%%%%%%%%
\section{The handbag mechanism}
\label{sec:handbag}
%%%%%%%%%%%%%%%%%%%%%%%%%%%%%%%%%%%%%%%%%%%%%%%%%%%%%%%%%%%%%%%%%%%

The arguments for factorization in the wide-angle region of electroproduction are the same as
for Compton scattering \ci{DFJK1} and photoproduction \ci{huang00}. Thus, we can restrict ourselves to
the repetition of the most important arguments for factorization and the phenomenological ingredients of the
handbag mechanism.
Prerequisite for factorization of the electroproduction amplitudes into hard subprocesses and soft form
factors, is that the Mandelstam variables $s$, $-t$ and $-u$ are much larger than $\Lambda^2$ where 
$\Lambda$ is a typical hadronic scale of order $1\,\gev$. The virtuality of the photon, $Q^2$, is not regarded
as a hard scale. Thus,
\be
Q^2 < -t, -u\,.
\label{eq:condition-1}
\ee
The mass of the pion is neglected. Corrections due to the nucleon mass, $m$,
of order $m^2/s$ or higher are also neglected.          
It is advantageous to work in
a symmetrical frame which is a center-of-mass frame (c.m.s.) rotated in such a way that the momenta of the
ingoing ($p$) and outgoing ($p'$) nucleons have the same light-cone plus components. In this frame the
skewness, defined by
\be
\xi\=\frac{(p-p')^+}{(p+p')^+}\,,
\ee
is zero. Starting from a c.m.s. in which the ingoing proton moves along the 3-direction whereas the
outgoing one is scattered by an angle $\theta$ in the 1-3 plane, one needs a rotation in
that plane by an angle $\vartheta$, defined by the condition
\be
|{\bf p}| (1+\cos{\vartheta}) \= |{\bf p'}|(1+\cos{(\vartheta-\theta)})  + {\cal O}(m^2/s)\,,
\label{eq:condition}
\ee
in order to arrive at the symmetric frame (${\bf p}$ and ${\bf p'}$ are the 3-momenta of the nucleons).
The solution of the condition \req{eq:condition}
is
\be
\cos{\vartheta}\=\frac{-1}{s+Q^2}\,\frac1{Q^4-4st}\,\Big[Q^2(Q^4+sQ^2-2st)+4st\sqrt{-us}\Big] + {\cal O}(m^2/s)\,.
\label{eq:varphi}
\ee
In the photoproduction limit, $Q^2\to 0$, this becomes
\be
\cos{\vartheta} \to \sqrt{\frac{-u}{s}} = \cos{\theta/2} + {\cal O}(m^2/s)\,.
\ee
In the new, rotated, c.m.s. the nucleon momenta read
\ba
p&=&\Big[p^+,\, \frac{(s+Q^2)^2}{8sp^+}\,\sin^2{\vartheta},\, - \frac{s+Q^2}{2\sqrt{s}}\,\sin{\vartheta},\, 0\Big]\,, \nn\\
p'&=&\Big[p^+,\, \frac{s}{8p^+} \sin^2{(\vartheta-\theta)},\, -\frac{\sqrt{s}}{2}\,\sin{(\vartheta-\theta)},\, 0\Big]\,,
\label{eq:symmetric-frame}
\ea
where 
\be
p^+\=- \frac{st}{Q^4-4st} \Big(\sqrt{s}+\sqrt{-u} +Q^2/(2\sqrt{s})\Big) + {\cal O}(m^2/s) \,.
\ee
The parameterization of the momenta of the virtual photon, $q$, and of the meson, $q'$, is obvious.   
As for wide-angle Compton scattering or photoproduction of mesons it is  assumed that the parton virtualities
are restricted by $k_i^2<\Lambda^2$ and that the intrinsic transverse parton momenta, $k_{\perp i}$, defined with
respect to their parent hadron's momentum, satisfy the condition $k^2_{\perp i}/x_i<\Lambda^2$. Here, $x_i$ denotes
the momentum fraction that parton $i$ carries. On
these premises one can show that the subprocess Mandelstam variables $\sh$ and $\uh$ coincide with those of
the full process, meson electroproduction off nucleons, up to 
corrections of order $\Lambda^2/s$,
\be
\th\=t\,, \quad \sh\=(k_j+q)^2\simeq (p+q)^2\=s\,, \quad 
\uh\=(k'_j-q)^2\simeq (p'-q)^2\=u\,,
\label{eq:sub-full}
\ee
where $k_j$ and $k_j'=k_j+q-q'$ denote the momenta of the active partons, i.e., the partons 
to which the photon couples. 
Thus, the active partons are approximately on-shell, move collinear with their parent hadrons and carry 
a momentum fraction close to unity, $x_j, x'_j\simeq 1$. As in deeply virtual exclusive scattering, the 
physical situation is that of a hard parton-level subprocess, $\gamma^* q_a\to P q_b$ ($P=\pi^\pm, \pi^0$),
and a soft emission and reabsorption of quarks from the nucleon. 
Up to corrections of order $\Lambda/\sqrt{-t}$ the light-cone helicity amplitudes for wide-angle
electroproduction of pions, ${\cal M}^P_{0\nu',\mu\nu}$, are then given
by a product of subprocess amplitudes, ${\cal H}$, which will be discussed in Sec.\ \ref{sec:subprocess},
and form factors, $R_i$ and $S_i$, which parameterize the soft physics that controls the emission from and
reabsorption of a quark by the nucleon~\footnote
     {$t_{min}$ is neglected in the wide-angle region.},
\ba
{\cal M}^P_{0+,\mu +}&=& \frac{e_0}{2}\, \sum_\lambda \Big[
              {\cal H}^P_{0\lambda,\mu\lambda}\, \left( R_V^{P}(t)
              +2\lambda  \, R_A^{P}(t) \right) \nn\\
        &&    - 2\lambda \, \frac{\sqrt{-t}}{2m} \,
{\cal H}^P_{0-\lambda,\mu\lambda}\,\bar{S}^{P}_T(t)\Big]\,, \nn\\
{\cal M}^P_{0-,\mu +} &=& \frac{e_0}{2}\, \sum_\lambda \Big[
                 \frac{\sqrt{-t}}{2m} \, {\cal H}^P_{0\lambda,\mu\lambda}\, R_T^{P}(t) \nn\\
           && -2\lambda \, \frac{t}{2m^2} \, {\cal H}^P_{0-\lambda,\mu\lambda}\,S^{P}_S(t)\Big]
                              + e_0 {\cal H}^P_{0-,\mu +}\,S^{P}_T(t)\,,
\label{eq:amplitudes}
\ea 
where $\mu$ denotes the helicity of the virtual photon and $\nu$ ($\nu'$) is the helicity
of the ingoing (outgoing) nucleon. The helicity of the active incoming quark is denoted
by $\lambda$ and $e_0$ is the positron charge. Note that, for the sake of legibility, helicities are labeled
by their signs only. 

The soft form factors, $R_i^{P}$ and $S_i^{P}$, represent specific flavor combinations of $1/x$-moments of
zero-skewness  GPDs. The form factors $R_V^P$, $R_A^P$ and $R_T^P$ are related to the helicity non-flip GPDs
$H$, $\widetilde{H}$ and $E$, respectively. 
The $S$-type form factors, $S^P_T$, $\bar{S}_T^P$ and $S^P_S$ are
related to the helicity-flip or transversity GPDs $H_T$, $\bar{E}_T$
and $\widetilde{H}_T$, respectively.
%\footnote{Note $\bar{E}_T=2\widetilde{H}_T+E_T$.}%
The GPDs $\tilde{E}$ and $\tilde{E}_T$ and their
associated form factors decouple at zero skewness.

The amplitudes for helicity configurations other than appearing in \req{eq:amplitudes} 
follow from parity invariance
\be
   {\cal M}^P_{0-\nu',-\mu-\nu} \= -(-1)^{\mu - \nu + \nu'} {\cal M}^P_{0\nu',\mu\nu}\,.
\label{eq:parity}   
\ee
An analogous relation holds for the subprocess amplitudes ${\cal H}$. With the help of \req{eq:parity}
the amplitudes for longitudinally polarized photons simplify drastically
\ba
   {\cal M}^P_{0+,0+}&=&e_0 {\cal H}^P_{0+,0+}\, R_A^P(t)\,, \nn\\
   {\cal M}^P_{0-,0+}&=& e_0 {\cal H}^P_{0-,0+}\, S^P_T(t)\,.
\label{eq:long-amplitudes}
\ea
   In contrast to the transverse amplitudes, twist-2 and twist-3 contributions are separated here: The
   nucleon helicity non-flip amplitude, ${\cal M}^P_{0+,0+}$, is pure twist-2 while the flip one is of
   twist-3 nature.~\footnote
   {In electroproduction of charged pions there is also a contribution from the pion pole. Even though the
     pole is important in the DVMP region it can be neglected in the wide-angle region where the pole at
     $t=m_\pi^2$ is very far away.}

The handbag mechanism can be interpreted as Feynman's end-point mechanism \ci{DFJK4}. As we said above the
handbag mechanism is restricted to the kinematical situation where all three Mandelstam variables are sufficiently
large compared to the hadronic scale $\Lambda^2$. It is, however, conceivable that in the asymptotic regime of
very large Mandelstam variables the hard perturbative mechanism \ci{BL80,ER80}, for which all partons inside the
nucleon participate in the hard process, dominates. Since each additional parton requires one more hard gluon to
be exchanged, the higher Fock components of the nucleon are suppressed by inverse powers of $s$ compared to the
valence Fock component. This is to be contrasted with the handbag mechanism where there is only one active parton.
All other partons inside the nucleon are spectators and the summation goes over all Fock components. 
Thus, it appears to be plausible that the handbag mechanism provides a larger contribution 
than the hard perturbative mechanism at least at large but not asymptotically large Mandelstam variables. 
Indeed the latter mechanism provides results for form
factors and WACS which are way below experiment if evaluated from plausible \da s. A calculation of wide-angle
photoproduction within that approach has only attempted once \ci{farrar91}. The results are at drastic variance
with experiment \ci{anderson76} and need verification since the applied integration method is questionable and
is known to fail for WACS \ci{brooks-dixon, vanderhaeghen-guichon}.

%%%%%%%%%%%%%%%%%%%%%%%%%%%%%%%%%%%%%%%%%%%%%%%%%%%%%%%%%%%%%%%%%%%
\section{The subprocess amplitudes}
\label{sec:subprocess}
%%%%%%%%%%%%%%%%%%%%%%%%%%%%%%%%%%%%%%%%%%%%%%%%%%%%%%%%%%%%%%%%%

\begin{figure}[t]
\begin{center}
\includegraphics[width=0.59\tw]{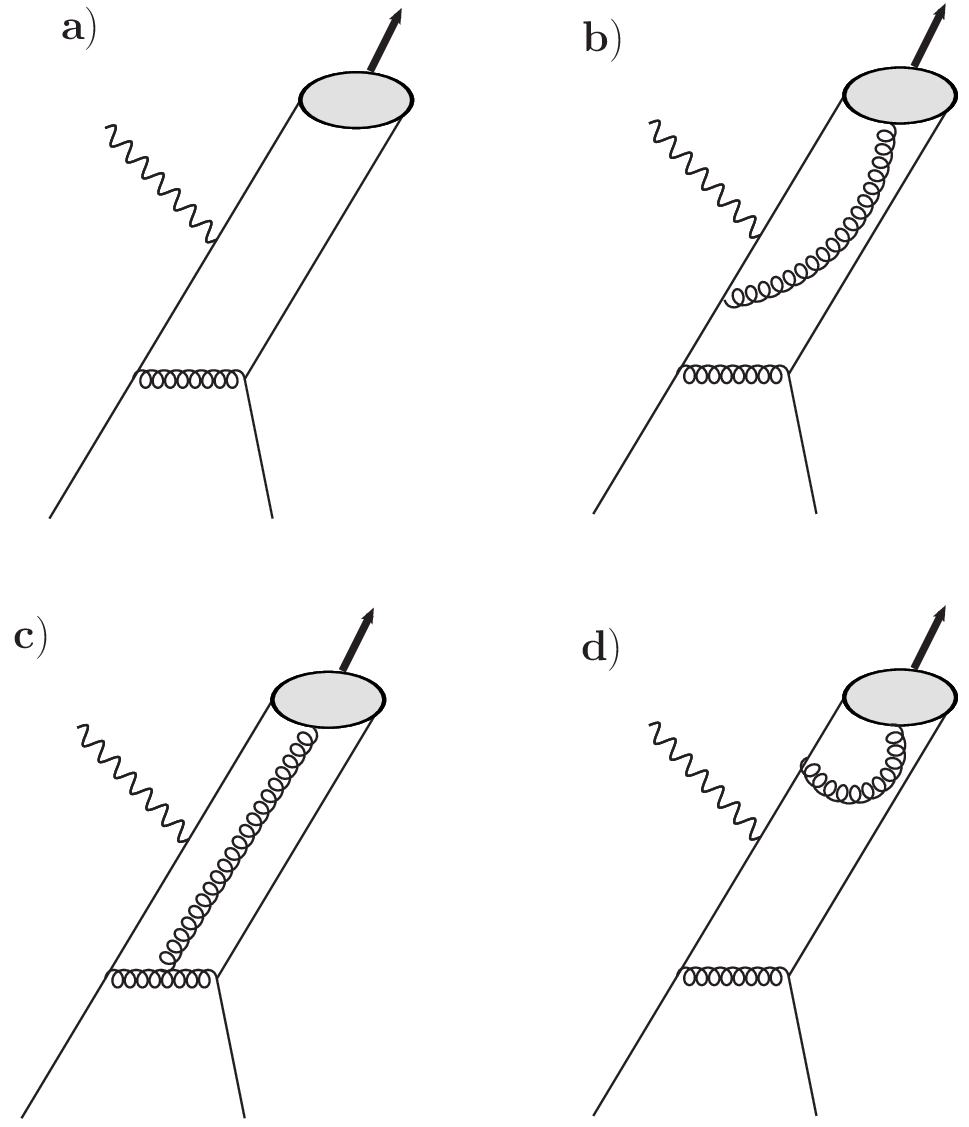}
\caption{\label{fig:graphs} Typical leading-order Feynman graphs for 
  $\gamma^* q\to P q$. a) for the 2-body twist-2 and twist-3 Fock components of the
  meson. b) and c) contributions from the $q\bar{q}g$ Fock component with and without 
triple gluon coupling. d) a soft contribution which is to be considered as a
part of the 2-body twist-3 \da s. }
\end{center}
\end{figure} 

The calculation of  the amplitudes for the subprocess $\gamma^* q_a\to P q_b$ to twist-3 accuracy
is a straightforward generalization of our calculation of the process  $\gamma q_a\to \pi^0 q_a$ \ci{KPK18}. 
In the definitions of the soft pion and nucleon matrix elements we are using light-cone gauge: the
plus component of the gluon field is zero. All possible Wilson lines become unity in that gauge.
It is to be stressed that our method for calculating the subprocess amplitudes is similar to the light-cone
collinear factorization approach advocated for in \ci{anikin02,wallon} for the case of electroproduction
of transversely polarized vector mesons.

Typical lowest-order Feynman graphs for the process of interest are depicted in 
Fig.\ \ref{fig:graphs}. The four graphs of type a) are relevant for the 2-body 
contributions. With the help of the projector of a massless $q\bar{q}$ pair on an outgoing massless pion 
\ci{signatures,beneke} ($P=\pi^\pm, \pi^0$, isospin invariance is assumed)
\ba
{\cal P}_{2,l m}^{P(ab)}&=&\frac{f_\pi}{2\sqrt{2N_C}}\,{\cal C}^{ab}_P\,\frac{\delta_{l m}}{\sqrt{N_C}}
                       \,\left\{\frac{\gamma_5}{\sqrt{2}}\,q\sla'\phiDA_\pi(\tau,\mu_F) 
                   + \mu_\pi(\mu_F)\frac{\gamma_5}{\sqrt{2}}\, \Big[\phiDA_{\pi p}(\tau,\mu_F)  \right. \nn \\ 
                 &&\left.  - \frac{i}{6}\sigma_{\mu\nu}\,\frac{q'{}^\mu k'{}_j^\nu}{q'\cdot k'_j}
                            \,\phiDA'_{\pi\sigma}(\tau,\mu_F)
                + \frac{i}{6} \sigma_{\mu\nu}\,q'{}^\mu\,\phiDA_{\pi\sigma}(\tau,\mu_F) 
                            \frac{\partial}{\partial k_{\perp\nu}} \Big] \right\}_{k_\perp\to 0}
\label{eq:projector-2}
\ea
the subprocess amplitudes for the twist-2 and the 2-body twist-3 contributions can easily be
calculated. The familiar twist-2  \da{} is denoted by $\phi_\pi$ whereas $\phi_{\pi p}$ and 
$\phi_{\pi\sigma}$ are the two 2-body twist-3 \da s. The \da{} $\phi'_{\pi\sigma}$ is the derivative of
$\phi_{\pi\sigma}$ with respect to the momentum fraction, $\tau$, the quark entering the meson carries.  
Their definitions can be found for instance  in App.\ A of \ci{KPK18}. For convenience we omit their
scale dependence (see App.\ A) in the list of variables in the following. Furthermore,  $f_\pi$
in \req{eq:projector-2} is the usual decay constant of the pion ($f_\pi=132\,\mev$); $a(l)$ and $b(m)$ represent
flavor (color) labels of the quark and antiquark, respectively.  $N_C$ is the number of colors.
The Dirac labels are omitted for convenience.  The flavor weight factors are
\ba
   {\cal C}^{uu}_{\pi^0}&=&-{\cal C}^{dd}_{\pi^0}\=\frac1{\sqrt{2}}\,, \qquad {\cal C}^{ud}_{\pi^+}\= {\cal C}^{du}_{\pi^-}\=1\,.
   \label{eq:flavor-weights}
   \ea
All other ${\cal C}^{ab}_P$ are zero.\\
In \req{eq:projector-2}, $k_{\perp}$ denotes the intrinsic transverse momentum of the quark entering
the meson. It is defined with respect to the meson's momentum, $q'$. 
The quark and antiquark momenta are thus given by
\be
k_q\=\tau q' +k_\perp\,, \qquad k_{\bar{q}}\=\taub q'-k_\perp
\ee
where $\taub=1-\tau$ and $q'\cdot k_\perp=0$. After the derivative in \req{eq:projector-2} is performed the collinear 
limit, $k_{\perp}=0$, is to be taken.

The mass parameter $\mu_\pi$ is large since it is the pion mass, $m_\pi$, enhanced
by the chiral condensate
\be
\mu_\pi\= \frac{m_\pi^2}{m_u+m_d}
\label{eq:mass-parameter}
\ee
by means of the divergence of the axial-vector current. Here, $m_{u(d)}$ are the current-quark masses
of the pion's constituents. Following \ci{GK5,GK6} we take the value $2\,\gev$ for this parameter
at the initial scale $\mu_0=2\gev$. Its uncertainty is, however, large~\footnote{
  Using, for instance, the values for the current-quark masses quoted in \ci{PDG}, one obtains
  $\mu_\pi(\mu_0)=2.64^{+0.11}_{-0.42}\,\gev$ for the case of the $\pi^0$. Values
  of $1.8\,\gev$ and $1.9\,\gev$ at $\mu_0$ are advocated for in \ci{huang} and 
    \ci{ball98}, respectively.}. 
The mass parameter evolves with the scale, see App.\ A.
For the factorization and renormalization scale we choose 
\be
\muR\=\muF\=\frac{\th \uh }{\sh}
\label{eq:muRmuF}
\ee
which takes care of the requirement that both $t$ and $u$ should be large. 
The strong coupling, $\als(\mu_R)$, is evaluated 
in the one-loop approximation
from $\Lambda_{\rm QCD}=0.22\,\gev$ and four flavors ($n_f=4$).

%%%%%%%%%%%%%%%%%%%%%%%%%%%%%%%%%%%%%%%%%%%%%%%%%%%%%%%%%%%%%%%%%%%%%%%%%%%%%%%%%%%%%%%%%%%%%%%%%%
\subsection{The twist-2 subprocess amplitudes}
\label{sec:twist-2}
%%%%%%%%%%%%%%%%%%%%%%%%%%%%%%%%%%%%%%%%%%%%%%%%%%%%%%%%%%%%%%%%%%%%%%%%%%%%%%%%%%%%%%%%%%%%%%%%%%%

As shown in \ci{huang00} at leading-order (LO) of perturbative QCD the twist-2 (light-cone) helicity
amplitude for transversely polarized photons reads
\ba
{\cal H}_{0\lambda,\,\mu\lambda}^{P,tw2} &=&  \frac12 \kappa_P^{(ab)} f_\pi 
                        \,\frac{\sqrt{-\th}}{\sh+Q^2}\,\int_0^1\,d\tau \,\phi_{\pi}(\tau) \nn \\
                                     & & \times \left[\left(1+2 \lambda \mu \right)
\left(
\frac{(\sh \tau+Q^2)(\sh+Q^2)-\uh Q^2 \taub}{\sh \taub  (Q^2 \taub-\th \tau)}
\, e_a
\right.
\right.
\nn \\
& & 
\left.\left.\hspace*{0.15\tw}
           + \frac{(\sh \tau-Q^2)(\sh+Q^2)-\uh Q^2 \taub}{\uh \tau (Q^2 \tau-\th \taub)} \,e_b \right)
             \right.\nn \\
& & \left. - \,\left(1-2 \lambda \mu \right) \left( \frac{\uh \, e_a}{(Q^2 \taub-\th \tau)}
            + \frac{\sh \taub \, e_b}{\tau(Q^2 \tau-\th \taub)} \right) \right]\, ,
\label{eq:Htw2T}
\ea
and for longitudinally polarized photons
\ba
{\cal H}_{0\lambda,\,0\lambda}^{P,tw2}&=&  2\sqrt{2} \lambda  \kappa_P^{(ab)} f_\pi 
                            \, \frac{Q \sqrt{-\uh \sh}}{\sh+Q^2}\, \int_0^1 \, d\tau\, \phi_{\pi}(\tau) \nn \\
& & \times \left(\frac{\uh \, e_a }{\sh (Q^2 \taub-\th \tau)} -
                \frac{(\th+\tau \uh) \, e_b }{\tau \uh (Q^2 \tau-\th \taub)}\right) 
\label{eq:Htw2L}
\ea
where
\be
\kappa_P^{(ab)}\=2\sqrt{2} \pi\,\als(\muR) \, {\cal C}^{ab}_P\,  \frac{C_F}{N_C}\,.
\ee
As usual, $C_F=(N_C^2-1)/(2 N_C)$ is a color factor and $e_a (e_b)$ is the charge of the flavor-$a (b)$ quark
in units of the positron charge. The summation over the same flavor labels is understood.
The twist-2 contribution only affects the subprocess amplitudes for quark helicity non-flip, the
quark helicity-flip amplitudes are zero. Leaving aside the pion \da{} the subprocess
amplitudes are expressed in terms of the Lorentz invariant Mandelstam variables. Thus, in any other c.m.s.
the expressions \req{eq:Htw2T} and \req{eq:Htw2L} hold too. 
In any case we assume that in the symmetric frame the pion \da{} is the usual one. 
%In any case we assume that in a c.m.s. in which factorization is shown, the pion \da{} is the usual one. 
%For the wide-angle case this is the symmetric
%frame \req{eq:symmetric-frame} but for the DVMP limit, $t\to 0$, which we occasionally discuss, this is Ji's
%frame \ci{ji98}.
  
In the photoproduction limit, $Q^2\to 0$, the longitudinal amplitude vanishes $\propto Q$
( see Tab.\ \ref{tab:power-behavior-s}). 
For the transverse amplitude,
taking into account the $\tau\to\taub$ symmetry property of pion DA,
we recover the result derived
in \ci{huang00}  which we explicitly quote for later use: 
\begin{eqnarray}
{\cal H}_{0\lambda,\,\mu\lambda}^{P,tw2} &\stackrel{Q^2\to 0}{\longrightarrow}&
        \frac12 \kappa_P^{(ab)}\,f_\pi\, \frac1{\sqrt{-\th}} \, \int_0^1 \, \frac{d\tau}{\tau}\, \phi_{\pi}(\tau)  \nn \\
 & \times & \left(\left(1+2 \lambda \mu \right) \sh
                      -\left(1-2 \lambda \mu \right) \uh \right)
\left(
\frac{e_a}{\sh} + \frac{e_b}{\uh} \right) \,.
\label{eq:Htw2Tphoto1}
\end{eqnarray}

\begin{table*}[t]
\caption{Power behavior of various twist-2 and twist-3 contributions for transverse and longitudinal photons
  for $Q^2\to 0$. The singularities \req{eq:Q-singularity-2} and \req{eq:Q-singularity-3} are already subtracted.
  Note that, according to the prerequisite of the handbag mechanism,
  $-\th$ and $-\uh$ are of the order of $\sh$.}
\renewcommand{\arraystretch}{2} 
\begin{center}
\begin{tabular}{| l || c | c| }
\hline
  \multicolumn{1}{|c||}{${\cal H}$}      & transverse  & longitudinal   \\[0.4em]
\hline   
twist-2     &   
$\displaystyle \sim \frac{f_\pi}{\sqrt{\sh}}$ & 
$\displaystyle \sim \frac{f_\pi}{\sqrt{\sh}}\, \frac{Q}{\sqrt{\sh}}$   \\[0.4em]
twist-3, 2-body, 
$\phi_{\pi p}$ & 
$\displaystyle \sim \frac{f_\pi\mu_\pi}{\sh}\,\frac{Q^2}{\sh}$ & 
$\displaystyle \sim \frac{f_\pi\mu_\pi}{\sh}\, \frac{Q}{\sqrt{\sh}}$ \\[0.4em]
twist-3, $C_F$, $\phi_{3\pi}$ & 
$\displaystyle \sim \frac{f_{3\pi}}{\sh}$ & 
$\displaystyle \sim \frac{f_{3\pi}}{\sh}\, \frac{Q}{\sqrt{\sh}}$  \\[0.4em]
twist-3, 3-body, $C_G$ & 
$\displaystyle \sim \frac{f_{3\pi}}{\sh}$ & 
$\displaystyle \sim \frac{f_{3\pi}}{\sh}\, \frac{Q}{\sqrt{\sh}}$ \\[0.4em]  
\hline
\end{tabular}
\end{center}
\label{tab:power-behavior-s}
\renewcommand{\arraystretch}{1.0}   
\end{table*} 

%%%%%%%%%%%%%%%%%%%%%%%%%%%%%%%%%%%%%%%%%%%%%%%%%%%%%%%%%%%%%%%%%%%%%%%%%%%%%%%%%%%%%%%%%%%%%%%%%%%%
\subsection{The 2-body twist-3 contributions}
\label{sec:2-body}
%%%%%%%%%%%%%%%%%%%%%%%%%%%%%%%%%%%%%%%%%%%%%%%%%%%%%%%%%%%%%%%%%%%%%%%%%%%%%%%%%%%%%%%%%%%%%%%%%%%%

Let us now turn to the 2-body twist-3 contribution to the subprocess amplitudes. It is tightly
connected to the 3-body twist-3 contribution through the equations of motion (EOMs). 
In fact, for light-cone gauge which we have chosen for the vacuum-pion matrix element (more details on that
choice for the twist-3 calculation have been given in \cite{KPK18}), 
the 2-body twist-3 \da s are related to the 3-body one, $\phi_{3\pi}(\tau_a,\tau_b,\tau_g)$,
integrated upon the momentum fraction the constituent gluon carries:
\ba
\tau \phi_{\pi p}(\tau) + \frac{\tau}{6}\, \phi'_{\pi\sigma}(\tau) - \frac1{3}\, \phi_{\pi \sigma}(\tau)
  & = & \phi^{EOM}_{\pi 2} (\taub)\,,  \nn\\
\bar{\tau} \phi_{\pi p}(\tau) - \frac{\taub}{6} \phi'_{\pi \sigma}(\tau) -\frac1{3} \phi_{\pi \sigma}(\tau)
& = & \phi^{EOM}_{\pi 2} (\tau)\,,
\label{eq:phieqmotion}
\ea
where
\be
\phi^{EOM}_{\pi 2}(\tau)\= 2 \frac{f_{3\pi}}{f_\pi\mu_\pi}\int_0^{\taub} \frac{d\tau_g}{\tau_g}\,
                                      \phi_{3\pi}(\tau,\taub-\tau_g,\tau_g)\,.
\label{eq:phi-EOM}
\ee
As the 2-body \da s $\phi_{3\pi}$ depends on the factorization scale, see App.\ A. We also
omit the scale dependence in its list of variables.
The parameter $f_{3\pi}$ plays a similar role as the pion decay constant for the twist-2 \da. 
It also evolves with the factorization scale.

With the help of the relation \req{eq:phieqmotion} for the 2-body twist-3 \da s 
we can express 
the corresponding helicity amplitudes in terms of, say, $\phi_{\pi p}$ and $\phi_{\pi 2}^{EOM} $:
\be
   {\cal H}^{P,tw3,q\bar{q}} \= {\cal H}^{P,\phi_{\pi p}} + {\cal H}^{P,\phi_{\pi 2}^{EOM}}\,.
\ee
Here the $\tau \to \taub$ symmetry property of $\phi_{\pi p}$ and $\phi_{\pi \sigma}$ was used.
Note that the 3-body \da{} possesses the symmetry property
\be
\phi_{3\pi}(\tau_a,\tau_b,\tau_g)\=\phi_{3\pi}(\tau_b,\tau_a,\tau_g)
\ee
which we also use in the following in order to simplify the expressions for the subprocess amplitudes.
For transversely polarized photons we find
\ba
{\cal H}_{0-\lambda,\,\mu\lambda}^{P,\phi_{\pi p}} 
      &=& \kappa_P^{(ab)}\,f_\pi \mu_\pi \, \frac{\sqrt{-\uh \sh}}{\sh+Q^2}
                       \,\frac{Q^2}{\th+Q^2}   \, \int_0^1 \, d\tau\, \frac{\phi_{\pi p}(\tau)}{Q^2 \taub-\th\tau}   \nn \\
      &\times & \left\{\frac{Q^2 \uh \left(2 \lambda + \mu \right) - \sh \th \left(2 \lambda - \mu \right) }{\sh} \right.   \nn \\
      & & \left.\times \left[\left(\frac{\th}{Q^2 \taub-\th \tau} - 1\right)
                           \left(\frac{e_a}{\sh} - \frac{e_b}{\uh} \right)
                + (\th+Q^2)\, \frac{e_b}{\uh^2} \right] \right. \nn \\
      & & \left. + \,4\mu\th\, \left( \frac{e_a}{\sh} - \frac{e_b}{\uh} \right) \right\}\, ,
\label{eq:Htw32partphipT}   
\ea
and
\ba
{\cal H}_{0-\lambda,\,\mu\lambda}^{P,\phi_{\pi 2}^{EOM}} 
      &=&  \frac12 \kappa_P^{(ab)}\,f_\pi \mu_\pi\,  \frac{\sqrt{-\uh \sh}}{\sh+Q^2}
                                 \frac{\th}{\th+Q^2}  \, \int_0^1 \, d\tau \, \phi_{\pi 2}^{EOM}(\tau) \nn \\
      &\times & \left\{\frac{Q^2 \uh \left(2 \lambda + \mu \right) -
\sh \th
\left(2 \lambda - \mu \right) 
}{\sh}
\right.
\nn \\
& &
\left.
\times
\left[
- \left(
\frac{ Q^2 + (Q^2 \taub - \th \tau)}{\taub \, (Q^2 \taub-\th \tau)^2}
+
\frac{\th - (Q^2 \tau-\th \taub)}{\tau \, (Q^2 \tau-\th \taub)^2}
\right)
\left(
\frac{e_a}{\sh}
-
\frac{e_b}{\uh}
\right)
\right.
\right.
\nn \\
& &
\left.
\left.
-
\left(
\frac{\tau}{\taub^2 \, (Q^2 \taub-\th \tau)}
+
\frac{1}{\tau \, (Q^2 \tau-\th \taub)}
\right)
\frac{\th+Q^2}{\uh}
\frac{e_b}{\uh}
\right]
\right.
\nn \\
& & 
\left.
- 2\mu \left(\frac{2 Q^2}{\taub \, (Q^2 \taub-\th \tau)} +
\frac{\th - Q^2}{\tau \, (Q^2 \tau-\th \taub)}
\right)
\left(
\frac{e_a}{\sh}
-
\frac{e_b}{\uh}
\right)
\right\}
\, .
\label{eq:Htw32partphiEOM2T}
\ea

For longitudinally polarized photons we obtain
\ba
{\cal H}_{0-\lambda,\,0\lambda}^{P,\phi_{\pi p}}
      &=&  -2\sqrt{2} \kappa_P^{(ab)} \,f_\pi \mu_\pi(\muF) 
\, \frac{Q \sqrt{-\th}}{\sh+Q^2}
\, \int_0^1 \, d\tau \frac{\phi_{\pi p}(\tau)}{Q^2 \taub-\th \tau} \nn \\
& & 
\times
\left( \frac{\th (Q^2+\sh \tau)}{Q^2 \taub-\th \tau} \left(
\frac{e_a}{\sh} - \frac{e_b}{\uh} \right)
- Q^2\,\frac{e_b}{\uh} \right) \, ,
\label{eq:Htw32partphipL}
\ea
and 
\ba
{\cal H}_{0-\lambda,\,0\lambda}^{P,\phi_{\pi 2}^{EOM}} &=&  
-\sqrt{2} \kappa_P^{(ab)} \,f_\pi \mu_\pi 
\, \frac{Q \sqrt{-\th}}{\sh+Q^2}
\, \int_0^1 
\, d\tau
\,  \phi_{\pi 2}^{EOM}(\tau) 
\nn \\
& \times &
\left[
\left(
\frac{Q^2 \uh}{(Q^2 \taub-\th \tau)^2}
+
\frac{\th \uh}{(Q^2 \tau-\th \taub)^2}
\right.
\right.
\nn \\
& &
\left.
\left.
-
\frac{\th}{\taub (Q^2 \taub-\th \tau)}
+
\frac{2 Q^2 \th - Q^2 \uh + \sh \th}{2\tau Q^2 (Q^2 \tau-\th \taub)}
\right)
\left(
\frac{e_a}{\sh}
-
\frac{e_b}{\uh}
\right)
\right.
\nn \\
& &
\left.
+
\left(
\frac{\th \tau}{\taub^2 (Q^2 \taub-\th \tau)}
+
\frac{\th}{\tau (Q^2 \tau-\th \taub)}
\right)
\frac{e_b}{\uh}
\right]
\, .
\label{eq:phi2L}
\ea

Because of the derivative with respect to the quark transverse momentum involved in the projector
\req{eq:projector-2} terms proportional to the square of propagators are generated in the amplitudes
\req{eq:Htw32partphipT} - \req{eq:phi2L}.
All these terms are unproblematic with one exception
\footnote{In this context it is important to realize that $\Phi_{\pi 2}^{EOM}$                                                                   vanishes at the end points $\tau=0$ and 1, see Eq. (59) below.}: 
the term $\propto Q^2\th/(\taub Q^2-\tau \th)^2$ in \req{eq:Htw32partphipT} leads to
\be
\lim_{Q^2\to 0} {\cal H}^{P,\phi_{\pi p}}_{0-\lambda,\mu\lambda}(Q^2)\= 
(2\lambda-\mu) \kappa_P^{(ab)}f_\pi \mu_\pi \sqrt{-\frac{\uh}{\sh}}\,
\left(\frac{e_a}{\sh}-\frac{e_b}{\uh}\right)\,\phi_{\pi p}(0)
\label{eq:photo-limit}
\ee
in the photoproduction limit, $Q^2\to 0$, provided the integration is performed before the limit is taken. 
Since the \da{} $\phi_{\pi p}$ does not vanish at $\tau =0$ the integral
\req{eq:photo-limit} is in conflict with our study of wide-angle pion photoproduction \ci{KPK18}
where we found that the $\phi_{\pi p}$ contribution to the subprocess amplitudes is zero. This latter result also
implies that the WW approximation is zero in wide-angle photoproduction.
We regard this apparent contradiction between photoproduction and the $Q^2\to 0$ limit of electroproduction as
a relic of the $k_\perp$-dependence of the projector \req{eq:projector-2}. Although the term \req{eq:photo-limit} is
of minor numerical importance we prefer to have a smooth transition from electro- to photoproduction which we
achieve by allowing for a mean-square quark transverse momentum, $\langle k_\perp^2\rangle$, in the propagator
$\taub Q^2-\tau \th$ of the relevant term
\be
\int_0^1 d\tau \phi_{\pi p}(\tau)\frac{Q^2 \th}{(Q^2\taub-\th\tau)^2} \longrightarrow
\int_0^1 d\tau \phi_{\pi p}(\tau)\frac{Q^2 \th}{(Q^2\taub-\th\tau + \langle k_\perp^2\rangle)^2}\,.
\ee
 With this regulator, being reminiscent to the $k_\perp$ dependence of the propagator, we arrive at
\be
\lim_{Q^2\to 0} {\cal H}^{P,\phi_{\pi p},reg}_{0-\lambda,\mu\lambda}(Q^2)\=0
\ee
and so obtain consistency with the photoproduction analysis. Following \ci{GK5} we adopt the value $0.5\,\gev^2$
for $\langle k_\perp^2\rangle$ in our numerical studies.
It is straightforward to show that the subprocess amplitude 
${\cal H}_{0-\lambda,0\lambda}^{P,\phi_{\pi p}}$,
Eq. \req{eq:Htw32partphipL}, vanishes in the photoproduction limit $Q^2\to 0$.

In Tab.\ \ref{tab:power-behavior-s} we list the power behavior of various contributions to the subprocess
amplitudes for $Q^2\to 0$. The $\phi_{\pi 2}^{EOM}$ part for longitudinally polarized photons, Eq.\ \req{eq:phi2L},
is singular in the photoproduction limit
\be
   {\cal H}^{P,\phi_{\pi 2}^{EOM}}_{0-\lambda,0\lambda} \stackrel{Q^2\to 0}{\longrightarrow} \frac1{\sqrt{2}} \kappa_P^{(ab)} 
                 f_\pi \mu_\pi\, \frac{\sqrt{-\th}}{Q} \int_0^1 \frac{d\tau}{\tau\taub} \phi_{\pi 2}^{EOM}(\tau)
                  \left(\frac{e_a}{\sh}-\frac{e_b}{\uh} \right)\,.
\label{eq:Q-singularity-2}
\ee                  

As we will see in the following section, this singularity is cancelled in the full subprocess amplitude by a
corresponding singularity in the 3-body contribution. Subtracting the singularity \req{eq:Q-singularity-2}
the remaining contribution to $ {\cal H}^{P, \phi_{\pi 2}^{EOM}}_{0-\lambda,0\lambda}$ vanishes in the photoproduction
limit, see Tab.\ \ref{tab:power-behavior-s}. Hence, the only 2-body twist-3 contribution that survives
in the photoproduction limit is  ${\cal H}^{P, \phi_2^{EOM}}_{0-\lambda,\mu\lambda}$ as we found in \ci{KPK18}.

%%%%%%%%%%%%%%%%%%%%%%%%%%%%%%%%%%%%%%%%%%%%%%%%%%%%%%%%%%%%%%%%%%%%%%%%%%%%%%%%%%%%%%
\subsection{The 3-body twist-3 contributions}
\label{sec:3-particle}
%%%%%%%%%%%%%%%%%%%%%%%%%%%%%%%%%%%%%%%%%%%%%%%%%%%%%%%%%%%%%%%%%%%%%%%%%%%%%%%%%%%%%%%

Typical lowest order Feynman graphs relevant for the 3-body twist-3 contributions are shown 
in Fig.\ \ref{fig:graphs} b-c. The 16 Feynman graphs (b) and (c) 
make up the 3-body contribution. Graphs of type (d) for which the constituent
gluon of the pion couples to one of its quark constituents, are  soft contributions and are to
be considered as parts of the 2-body twist-3 meson \da. In light-cone gauge
the gluon field, $A_\mu(x)$, appearing in perturbation theory, is related to the gluon field
strength tensor $G_{\mu\nu}$ which defines the 3-body \da{}, $\phi_{3\pi}$, \ci{kogut-soper}
by ($n$ being a light-like vector)
\be 
A^r_\rho(z) \=\lim_{\eps\to 0} n^\nu \int_0^\infty d\sigma e^{-\eps\sigma} 
                                              G^r_{\rho\nu}(z+ n\sigma)\,.
\label{eq:A-G}
\ee
In connection with the definition of the 3-body twist-3 \da{} (see, for instance, \ci{ball98,braun90})
this relation between $A_{\mu}$ and $G_{\mu\nu}$ allows to derive the 3-body projector, $q\bar{q}g\to \pi$, 
to be used in perturbative calculations involving the $q\bar{q}g$ Fock component \ci{KPK18}.
For an outgoing pion this projector reads
\be
{\cal P}_{3,l m}^{P(ab)\rho,r}\=
             \frac{-i}{g}\,\frac{f_{3\pi}(\mu_F)}{2 \sqrt{2 N_C}}
\,{\cal C}^{ab}_P \, \frac{\left(t^r\right)_{l m}}{C_F\sqrt{N_C}}\, 
           \frac{\gamma_5}{\sqrt{2}} \, \sigma_{\mu\nu} q'{}^{\mu} g_\perp^{\nu\rho}\, 
  \frac{\phiDA_{3\pi}(\tau_a,\tau_b,\tau_g, \mu_F)}{\tau_g} 
\,. 
\label{eq:projector-3}
\ee
The transverse metric tensor is defined as
\be
g^{\nu\rho}_\perp\= \left(g^{\nu\rho}-\frac{k_j'{}^\nu q'{}^\rho 
                                     + q'{}^\nu k_j'{}^\rho}{k'_j\cdot q'}
                          \right)\,,
\label{eq:transverse-metric-tensor}
\ee
and $t^r=\lambda^r/2$ is the SU(3) color matrix for a gluon of color $r$ while $g$ denotes the QCD coupling. 

For the sake of legibility we split the 3-body contribution into two parts:
\be
   {\cal H} ^{P, tw3,q\bar{q}g}\= {\cal H}^{P,q\bar{q}g, C_G} + {\cal H}^{P,q\bar{q}g, C_F}\,.
\ee
The first part is proportional to the color factor
\be
C_G\=C_F-\frac12 C_A
\, ,
\ee
which appears only in the 3-body twist-3 contributions. Here, as usual, $C_A=N_C$. 
The second part, ${\cal H}^{tw3,q\bar{q}g,C_F}$, is proportional to $C_F$ as is the case for the
2-body contributions. For transversely polarized photons the 3-body twist-3 contributions read
\ba
{\cal H}_{0-\lambda,\,\mu\lambda}^{P,q\bar{q}g, C_G} &=&  \kappa_P^{(ab)}\,f_{3\pi} 
\, \frac{C_G}{C_F} 
\, \frac{\sqrt{-\uh \sh}}{\sh+Q^2}
\nn \\
&  \times&
\big(
Q^2 \uh
\left(2 \lambda + \mu \right)
-
\sh \th 
\left(2 \lambda - \mu \right)
\big)
\left(
\frac{e_a}{\sh^2 \uh}
+
\frac{e_b}{\sh \uh^2}
\right)
\nn \\
&\times&
\, \int_0^1 
\, \frac{d\tau}{\taub}
\, \int_0^{\taub} 
\, \frac{d\tau_g}{\tau_g}
\, \phi_{3\pi}(\tau,\taub-\tau_g,\tau_g)
\nn\\
& \times & 
\left(
\frac{1}{\taub-\tau_g}
-\frac{\th}{(Q^2 \taub-\th \tau) \tau_g }
\right)\,,
\label{eq:Htw3CGT}
\ea
and 
\ba
{\cal H}_{0-\lambda,\,\mu\lambda}^{P,q\bar{q}g, C_F} &=&  \kappa_P^{(ab)} \,f_{3\pi}
\, \frac{\sqrt{-\uh \sh}}{\sh+Q^2}
 \int_0^1 
\, d\tau
\, \int_0^{\taub} 
\, \frac{d\tau_g}{\tau_g} \, \phi_{3\pi}(\tau,\taub-\tau_g,\tau_g)
\nn \\
& \times&
\left\{
\frac{
Q^2 \uh
\left(2 \lambda + \mu \right) 
-
\sh \th
\left(2 \lambda - \mu \right) 
}{\sh}
\right. \nn \\
&\times &  \left.
\left[
\left(
\frac{e_a}{\sh} 
-
\frac{e_b}{\uh}
\right)
\left(
\frac{\tau_g}{\taub (Q^2 \taub-\th \tau) (\taub-\tau_g)}
\right. \right. \right. \nn \\
&+  & \left. \left. \left.
\frac{\th}{\tau (Q^2 \tau-\th \taub)(Q^2 (\tau+\tau_g)-\th (\taub-\tau_g))}
\right)
\right.
\right.
\nn \\
&- &
\left.
\left.  \hspace*{-0.03\tw}
  \frac{Q^2 \tau_g-\th (1-\tau_g)}{\taub (Q^2 \taub-\th \tau) (\taub-\tau_g)}
\frac{e_b}{\uh^2}
\right]
+\, \frac{1}{\tau} \,
\frac{2\mu\th}{Q^2 \tau-\th \taub}
\left(
\frac{e_a}{\sh} 
-
\frac{e_b}{\uh}
\right)
\right\}
\, ,
\quad
\label{eq:Htw3CFT}
\ea
respectively. For longitudinally polarized photons they are given by
\ba
{\cal H}_{0-\lambda,\,0\lambda}^{P,q\bar{q}g, C_G} &=& 2\sqrt{2}\, \kappa_P^{(ab)} \,f_{3\pi}\, 
\,\frac{C_G}{C_F} 
\, \frac{Q \sqrt{-\th} }{\sh+Q^2}
\left(
\frac{e_a}{\sh}
+
\frac{e_b}{\uh}
\right)
\nonumber \\
& & \times
\, \int_0^1 
\, \frac{d\tau}{\taub}
\, \int_0^{\taub} 
\, \frac{d\tau_g}{\tau_g}
\, \phi_{3\pi}(\tau,\taub-\tau_g,\tau_g)
\nonumber \\
& & \times
\left(
\frac{1}{\taub-\tau_g}
-\frac{\th}{(Q^2 \taub-\th \tau) \tau_g }
\right)
\, ,
\label{eq:Htw3CGL}
\end{eqnarray}
and
\ba
{\cal H}_{0-\lambda,\,0\lambda}^{P,q\bar{q}g, C_F} &=&   2\sqrt{2}\,\kappa_P^{(ab)} \,f_{3\pi}\,
\frac{ Q \sqrt{-\th} }{\sh+Q^2}
 \, \int_0^1 
\, d\tau
\, \int_0^{\taub} 
\, \frac{d\tau_g}{\tau_g}
\nonumber \\
&   \times&
\, \phi_{3\pi}(\tau,\taub-\tau_g,\tau_g)
\left\{
\left(
\frac{e_a}{\sh}
-
\frac{e_b}{\uh}
\right)
\left(
\frac{\uh \tau_g }{\taub(Q^2 \taub-\th \tau)(\taub-\tau_g)}
\right.\right. \nonumber \\
& & \left. \left.
+
\frac{\th \uh }{\tau (Q^2 \tau-\th \taub) (Q^2 (\tau+\tau_g)-\th (\taub-\tau_g))}
\right.
\right.
\nonumber \\
& & 
\left.
\left.
+\;\frac{Q^2 \uh + \sh \th}{2 Q^2 \tau(Q^2 \tau-\th \taub)}
\right)
+ \; \frac{\th- (\th+Q^2) \tau_g}{\taub (Q^2 \taub-\tau \th) (\taub-\tau_g) }
\frac{e_b}{\uh}
\right\}\,.
\label{eq:Htw3CFL}
\end{eqnarray}

Evidently, ${\cal H}^{P,q\bar{q}g, C_F}_{0-\lambda,0\lambda}$ is singular in the photoproduction limit
\ba
   {\cal H}^{P,q\bar{q}g, C_F}_{0-\lambda,0\lambda}&\stackrel{Q^2\to 0}{\longrightarrow}&
   -\sqrt{2}\,\kappa_P^{(ab)} \,f_{3\pi}\, \frac{\sqrt{-\th}}{Q}\,
     \left(\frac{e_a}{\sh}-\frac{e_b}{\uh}\right) \nn\\
     &\times& \int_0^1\frac{d\tau}{\tau\taub} \int_0^{\taub} \frac{d\tau_g}{\tau_g}
     \phi_{3\pi}(\tau,\taub-\tau_g,\tau_g)\,.
  \label{eq:Q-singularity-3}
\ea
The $\tau_g$-integral in this relation is just $f_\pi\mu_\pi\phi_{\pi 2}^{EOM}/(2 f_{3\pi})$, see \req{eq:phi-EOM}.
Using this we notice that the singular part of $ {\cal H}^{P,q\bar{q}g, C_F}_{0-\lambda,0\lambda}$ 
exactly cancels that of ${\cal H}^{P,\phi_{\pi 2}^{EOM}}_{0-\lambda,0\lambda}$ given in \req{eq:Q-singularity-2}. 
Subtracting this singularity from ${\cal H}^{P,q\bar{q}g,C_F}_{0-\lambda,0\lambda}$ this amplitude vanishes proportional 
to  $Q$ in the photoproduction limit
as ${\cal H}^{P,q\bar{q}g, C_G}_{0-\lambda,0\lambda}$ does, see Tab.\ \ref{tab:power-behavior-s}. 

%%%%%%%%%%%%%%%%%%%%%%%%%%%%%%%%%%%%%%%%%%%%%%%%%%%%%%%%%%%%%%%%%%%%%%%%%%%%%%%%%%%%%%
\subsection{Remarks concerning gauge invariance}
%%%%%%%%%%%%%%%%%%%%%%%%%%%%%%%%%%%%%%%%%%%%%%%%%%%%%%%%%%%%%%%%%%%%%%%%%%%%%%%%%%%%%%
Probing current conservation of the twist-3 amplitudes we find that the 2-body twist-3 contributions
proportional to $\phi_{\pi p}$
as well as the 3-body twist-3 contributions proportional to $C_G$ respect it while the other two don't.
Replacing the photon polarization vector by the corresponding momentum in the $\phi_{\pi 2}^{EOM}$
contribution the term
\be
   {\cal H}_{elm}^{P,\phi_{\pi 2}^{EOM}}\=
\frac{\kappa_P^{(ab)}}{\sqrt{2}} f_\pi\mu_\pi\,\sqrt{-\th}\,(\th+Q^2)\,
     \left(\frac{e_a}{\sh} - \frac{e_b}{\uh}\right)
                 \int_0^1 \frac{d\tau}{\tau(\taub\th -\tau Q^2)}\, \phi_{\pi 2}^{EOM}
\label{eq:current-cons}
\ee
%is left over which is cancelled by a corresponding term from the 3-body twist-3 contributions
is left over. Expressing $\Phi_{\pi 2}^{EOM}$ through the 3-body \da\
 with the help of Eq. \req{eq:phi-EOM} one sees that \req{eq:current-cons} 
is cancelled by a corresponding term from the 3-body twist-3 contributions
proportional to  $C_F$.
Hence, the sum of ${\cal H}^{P,\phi_2^{EOM}}$ and ${\cal H}^{P,q\bar{q}g,CF}$ respects current
conservation.

We also have proven that the 2- and 3-body twist-3 contributions 
are separately gauge invariant with respect to the choice of the covariant gluon propagator
\be
g^{\mu\nu} - \zeta_c\,\frac{k_g^\mu\,k_g^\nu}{k_g^2}
\ee
where $k_g$ is the momentum of the virtual gluon and $\zeta_g$ an arbitrary gauge parameter.
We also checked the invariance of the subprocess amplitudes with respect to axial light-cone gauges
\ci{lee,landshoff}  
\be
g^{\mu\nu} - \, \frac{n^\mu k_g^\nu + n^\nu k_g^\mu}{n\cdot k_g}
\, .
\ee
where $n$ is an arbitrary light-like vector and $n\cdot k_g\neq 0$.
In contrast to the covariant gauges the separate pieces, ${\cal H}^{P,\phi_{\pi 2}^{EOM}}$ and ${\cal H}^{P,q\bar{q}g,C_F}$,
are not invariant with respect to axial gauges but their sum does not depend on $n$.
Thus, we can conclude that the twist-3 subprocess amplitudes are QCD gauge invariant.
The choice of gauge for the external gluon was discussed in \ci{KPK18}.
%%%%%%%%%%%%%%%%%%%%%%%%%%%%%%%%%%%%%%%%%%%%%%%%%%%%%%%%%%%%%%%%%%%%%%%%%%%%%%%%%%%%%%
\subsection{The complete twist-3 amplitude}
\label{sec:twist-3}
%%%%%%%%%%%%%%%%%%%%%%%%%%%%%%%%%%%%%%%%%%%%%%%%%%%%%%%%%%%%%%%%%%%%%%%%%%%%%%%%%%%%%%%
In order to present finally a gauge invariant result we have to add ${\cal H}^{P,\phi_{\pi 2}^{EOM}}$
(\req{eq:Htw32partphiEOM2T}, \req{eq:phi2L}) and ${\cal H}^{P,q\bar{q}g,C_F}$ (\req{eq:Htw3CFT}, \req{eq:Htw3CFL})
\begin{equation}
{\cal H}^{P,C_F,\phi_{3\pi}}\= {\cal H}^{P,\phi_{2}^{EOM}} + {\cal H}^{P,q\bar{q}g, C_F} \, .
\label{eq:phi3Psum}
\end{equation}
Since $\phi_{\pi 2}^{EOM}$ is an integral upon the 3-body \da{} $\phi_{3\pi}$, see Eq.\ \req{eq:phi-EOM},
the sum \req{eq:phi3Psum} can solely be expressed through $\phi_{3\pi}$ and simplifications occur.

For transversely polarized photons the sum \req{eq:phi3Psum} reads
\begin{eqnarray}
{\cal H}_{0-\lambda,\,\mu\lambda}^{P,C_F, \phi_{3\pi}} &=&  \kappa_P^{(ab)}\, f_{3\pi}
\, \frac{\sqrt{-\uh \sh}}{\sh+Q^2}
\, \int_0^1 
\, d\tau
\, \int_0^{\taub} 
\, d\tau_g
\, \phi_{3\pi}(\tau,\taub-\tau_g,\tau_g)
\nonumber \\[0.01\tw]
& \times&
\left\{
\frac{
Q^2 \uh
\left(2 \lambda + \mu \right) 
-
\sh \th
\left(2 \lambda - \mu \right) 
}{\sh}
\left[
\left(
\frac{e_a}{\sh} 
-
\frac{e_b}{\uh}
\right)
\right.
\right.
\nonumber \\[0.01\tw]
& \times &
\left.
\left.
\left(
\frac{1}{\taub (Q^2 \taub-\th \tau) (\taub-\tau_g)} \right.\right.\right.\nn\\[0.01\tw]
&&\left.\left.\left.
- \frac{\th (\th+Q^2)}{\tau (Q^2 \tau-\th \taub)^2
(Q^2 (\tau+\tau_g)-\th (\taub-\tau_g))}
\right.
\right.
\right.
\nonumber \\[0.01\tw]
& &
\left.
\left.
\left.
-
\frac{Q^4 \, \th\, (\th -Q^2) (\taub-\tau) }{(\th +Q^2) \,\tau \taub 
(Q^2 \taub-\th \tau)^2 (Q^2 \tau-\th \taub)^2 \tau_g}
\right)
\right.
\right.
\nonumber \\[0.01\tw]
& &
\left.
\left.
- 
\left(
\frac{Q^2\,  \th \, (\taub-\tau)}{\tau \taub 
(Q^2 \taub-\th \tau)(Q^2 \tau-\th \taub) \tau_g}
+
\frac{1}{\taub^2 (\taub-\tau_g)}
\right)
\frac{e_b}{\uh^2}
\right]
\right.
\nonumber \\[0.01\tw]
&+ &
\left.
2 \mu
\left(
\frac{e_a}{\sh} 
-
\frac{e_b}{\uh}
\right)
\frac{2 Q^4 \, \th \, (\taub-\tau)}{(\th +Q^2) \,\tau \taub 
(Q^2 \taub-\th \tau) (Q^2 \tau-\th \taub) \tau_g}
\right\}
\, .
\label{eq:Htw3CFphi3P}
\end{eqnarray}
For longitudinally polarized photons we have
\begin{eqnarray}
 {\cal H}_{0-\lambda,\,0\lambda}^{P,C_F, \phi_{3\pi}}    &=&  2\sqrt{2}\, \kappa_P^{(ab)}\,f_{3\pi}
\, \frac{Q \sqrt{-\th}}{\sh+Q^2}
  \, \int_0^1 
\, d\tau
\, \int_0^{\taub} 
\, d\tau_g
\, \phi_{3\pi}(\tau,\taub-\tau_g,\tau_g)
\nonumber \\[0.01\tw]
&  \times &
\left\{
\left(
-
\frac{Q^2 \uh }{(Q^2 \taub-\th \tau)^2 \tau_g}
-
\frac{\th (Q^2 \tau-\th \taub)-Q^2 \uh \tau}{\tau(Q^2 \tau-\th \taub)^2 \tau_g} \right.\right.\nn\\[0.01\tw]
&+& \left.\left.  \qquad
\frac{\th (\taub-\tau_g)+\uh \tau_g}{\taub (Q^2 \taub-\th \tau)\tau_g(\taub-\tau_g)}
\right.
\right.
\nonumber \\[0.01\tw]
&- & 
\left.
\left.
\frac{\th \uh (\th+Q^2)}{\tau (Q^2 \tau-\th \taub)^2
(Q^2 (\tau+\tau_g)-\th (\taub-\tau_g))}
\right)
\left(
\frac{e_a}{\sh}
-
\frac{e_b}{\uh}
\right)
\right.
\nonumber \\[0.01\tw]
&- & \left.\left(
\frac{Q^2\,\th \,(\taub-\tau)}{\tau \taub  
(Q^2 \tau-\th \taub)
(Q^2 \taub-\tau \th)\, \tau_g}
+
\frac{1}{\taub^2 (\taub- \tau_g)}
\right)
\frac{e_b}{\uh}
\right\}
\, .
\label{eq:Htw3CFLphi3P}
\end{eqnarray}

There is no singularity for $Q^2\to 0$
in the amplitude for longitudinally polarized photons, the singularities \req{eq:Q-singularity-2}
and \req{eq:Q-singularity-3} cancel as mentioned before. The amplitude ${\cal H}^{P,C_F,\phi_{3\pi}}_{0-\lambda, 0\lambda}$
vanishes proportional to $Q$ for $Q^2\to 0$. 

The complete twist-3 subprocess amplitude is given by 
\be
   {\cal H}^{P,tw3}\= {\cal H}^{P,\phi_{\pi p},reg}  +  {\cal H}^{P,q\bar{q}g, C_G} + {\cal H}^{P,C_F,\phi_{3\pi}}\,.
\label{eq:tw3-subamp}
\ee
It meets all theoretical requirements concerning gauge invariance. For longitudinally polarized photons
the three contributions in Eq.\ \req{eq:tw3-subamp} vanish in the photoproduction limit as well as in the DVMP
limit, see Tab.\ \ref{tab:power-behavior-s} and \ref{tab:power-behavior-Q}. 

\begin{figure}[t]
\begin{center}
\includegraphics[width=0.4\tw]{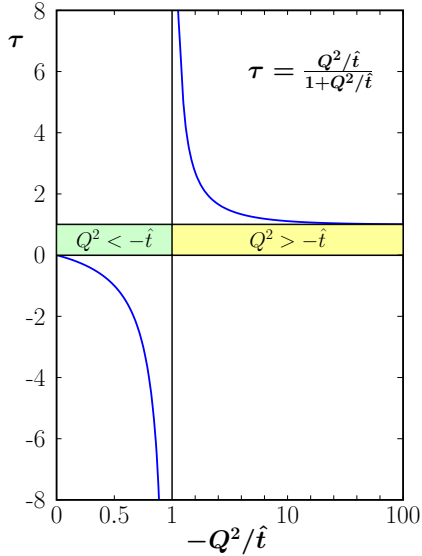}
\caption{\label{fig:pole} The function $\tau=\frac{Q^2/\th}{1+Q^2/\th}$. For $-Q^2/\th>1$ a logarithmic scale is used.}
\end{center}
\end{figure}

For non-zero $\th$ and $Q^2$ the propagator denominators $\tau Q^2-\taub\th$ and $\taub Q^2-\tau\th$ don't have a
pole in the physical region $0\leq \tau\leq 1$, see Fig.\ \ref{fig:pole}. Consequently,
all twist-3 contributions are free from singularities generated in the end-point regions
where either $\tau$ or $\taub$ tends to zero. Possible factors $1/\tau$ or $1/\taub$ are cancelled by the end-point
zeros of $\phi_{3\pi}$, i.e.\  $\phi_{\pi 2}^{EOM}$. In the limits $\th\to 0$ and $Q^2\to 0$, however, propagator poles 
occur which lead to singularities in the $\phi_{\pi p}$ contribution since this \da{} does not vanish for $\tau\to 0$ or 1.
For the photoproduction limit the singularity has already been discussed by us, see Eq.\ \req{eq:photo-limit}
and the ensuing paragraph. For the DVMP limit see section \ref{sec:dvmp}.

%%%%%%%%%%%%%%%%%%%%%%%%%%%%%%%%%%%%%%%%%%%%%%%%%%%%%%%%%%%%%%%%%%%%%%%%%%%%%%%%%%%%%%%%%%%%%%
\subsection{The photoproduction limit, $Q^2\to 0$}
%%%%%%%%%%%%%%%%%%%%%%%%%%%%%%%%%%%%%%%%%%%%%%%%%%%%%%%%%%%%%%%%%%%%%%%%%%%%%%%%%%%%%%%%%%%%%%
For later use we also quote the complete photoproduction limit of the twist-3 subprocess amplitude
\begin{eqnarray}
{\cal H}_{0-\lambda,\,\mu\lambda}^{P,tw3} &\stackrel{Q^2\to 0}{\longrightarrow}& \kappa_P^{(ab)}\,f_{3\pi}\, 
\left(2 \lambda - \mu \right) \, \sqrt{-\uh \sh}  \nn \\
& \times &
 \int_0^1 
\, d\tau
\, \int_0^{\taub} 
\, \frac{d\tau_g}{\tau_g}
\, \phi_{3\pi}(\tau,\taub-\tau_g,\tau_g)
\nonumber \\[0.1cm]
&  \times &
\left[ \left(
\frac{1}{\taub^2 }
-
\frac{1}{\taub(\taub-\tau_g)}
\right)
\left(
\frac{e_a}{\sh^2 }
+
\frac{e_b}{\uh^2}
\right)
\right.
+
\nonumber \\
&- & 
\left.
\frac{C_G}{C_F} \,
\,\frac{2}{\tau \tau_g}\, \frac{\th}{\sh\uh}
\left( \frac{e_a}{\sh} + \frac{e_b}{\uh} \right)
\right]
\,.
\label{eq:Htw3Tphoto}
\end{eqnarray}
This amplitude is in agreement with the results presented in \ci{KPK18}. 
It is free of end-point singularities.
The $Q^2\to 0$ limit of the twist-2 subprocess amplitude is given in Eq.\ \req{eq:Htw2Tphoto1}. 

Properties of the photoproduction limit have been discussed above. The behavior of the various
contributions to the subprocess for $Q^2\to 0$ are listed in Tab.\ \ref{tab:power-behavior-s}. 

%%%%%%%%%%%%%%%%%%%%%%%%%%%%%%%%%%%%%%%%%%%%%%%%%%%%%%%%%%%%%%%%%%%%%%%%%%%%%%%%%%%%%%%%%%%%%%%
\subsection{The DVMP limit, $\th\to 0$}
\label{sec:dvmp}
%%%%%%%%%%%%%%%%%%%%%%%%%%%%%%%%%%%%%%%%%%%%%%%%%%%%%%%%%%%%%%%%%%%%%%%%%%%%%%%%%%%%%%%%%%%%%%
The subprocess amplitudes discussed in this section hold in any c.m.s. as well as in the limits
$Q^2\to 0$ and $\th\to 0$. The latter limit is the region of DVMP
where factorization of the leading-twist contribution has been shown to hold for large $Q^2$ \ci{collins},
for instance in Ji's frame \ci{ji98}. In this frame in which skewness is non-zero, $\sh$ and $\uh$
are related to $Q^2$ by
\be
\sh\= \frac{x-\xi}{2\xi}\,Q^2\,, \qquad \uh\=-\frac{x+\xi}{2\xi}\, Q^2\,,
\ee
i.e.\ $\sh$ and $\uh$ are of order of $Q^2$. The behavior of the various contributions to the subprocess
amplitudes for $\th\to 0$ are compiled in Tab.\ \ref{tab:power-behavior-Q}. One sees from that table that
the asymptotically dominant contribution comes from longitudinally polarized photons at the twist-2 level.
In fact, the $\th\to 0$ limit
of \req{eq:Htw2L} is the familiar LO, leading-twist result \ci{collins}. On the other hand,
the transverse leading-twist amplitude vanishes proportional to $\sqrt{-\th}$ and, compared to the
longitudinal amplitude, is suppressed by $1/Q$. Table \ref{tab:power-behavior-Q} also reveals that all twist-3
contributions for longitudinally polarized photons vanish $\sim \sqrt{-\th}$ as a consequence of angular
momentum conservation. The twist-3 contributions for transversely polarized photons remain finite for $\th=0$
but are suppressed by $\mu_\pi/Q$ or $f_{3\pi}/Q$ compared to the dominant twist-2 contribution for longitudinally
polarized photons.

\begin{table*}[t]
\caption{Power behavior of various twist-2 and twist-3 contributions for transverse and longitudinal photons
  for $\th\to 0$ in the generalized Bjorken regime where $\sh$ and $-\uh$ are of the order of $Q^2$.}
\label{tab:power-behavior-Q}
\renewcommand{\arraystretch}{2} 
\begin{center}
\begin{tabular}{| l || c | c| }
\hline   
 \multicolumn{1}{|c||}{${\cal H}$}  & transverse  & longitudinal   \\[0.4em]
\hline   
twist-2     &   
$\displaystyle \sim \frac{f_\pi}{Q}\, \frac{\sqrt{-\th}}{Q}$ & 
$\displaystyle \sim \frac{f_\pi}{Q}$   \\[0.2em]
twist-3, 2-body, $\phi_{\pi p}$ & 
$\displaystyle \sim \frac{f_\pi\mu_\pi}{Q^2}$ & 
$\displaystyle \sim \frac{f_\pi\mu_\pi}{Q^2}\, \frac{\sqrt{-\th}}{Q}$ \\[0.4em]
twist-3, $C_F$, $\phi_{3\pi}$ & 
$\displaystyle \sim \frac{f_{3\pi}}{Q^2}$ & 
$\displaystyle \sim \frac{f_{3\pi}}{Q^2}\, \frac{\sqrt{-\th}}{Q}$  \\[0.4em]
twist-3, 3-body, $C_G$ & 
$\displaystyle \sim \frac{f_{3\pi}}{Q^2}$ & 
$\displaystyle \sim \frac{f_{3\pi}}{Q^2}\, \frac{\sqrt{-\th}}{Q}$ \\[0.4em]  
\hline
\end{tabular}
\end{center}
\renewcommand{\arraystretch}{1.0}   
\end{table*}

It is also informative to see the twist-3 subprocess amplitudes in the DVMP limit:
\ba
   {\cal H}^{P,\phi_{\pi p}}_{0-\lambda,\mu\lambda}&\stackrel{\th\to 0}{\longrightarrow}&
        (2\lambda+\mu)\, \kappa_P^{(ab)}\,f_\pi\mu_\pi \,\sqrt{-\frac{\uh}{\sh}}\,
                         \Big[ \frac{e_a}{\sh} + \frac{\sh}{\uh} \frac{e_b}{\uh}\Big]\,
                         \int_0^1 \frac{d\tau}{\taub} \phi_{\pi p}(\tau)\,, \nn\\
{\cal H}^{P,q\bar{q}g, C_G}_{0-\lambda,\mu\lambda} &\stackrel{\th\to 0}{\longrightarrow}&
                (2\lambda + \mu)\, \kappa_P^{(ab)}\,f_{3\pi}\, \big(1-\frac12\,\frac{C_A}{C_F}\big)
          \frac{Q^2}{\sqrt{-\sh\uh}} \left(\frac{e_a}{\sh} + \frac{e_b}{\uh}\right) \nn\\
          && \times\; \int_0^1 \frac{d\tau}{\taub} \, \int_0^{\taub} \frac{d\tau_g}{\tau_g (\taub-\tau_g)}\,
          \phi_{3\pi}(\tau,\taub-\tau_g,\tau_g)\,, \nn\\
    {\cal H}^{P, C_F,\phi_{3\pi}}_{0-\lambda,\mu\lambda} &\stackrel{\th\to 0}{\longrightarrow}&
                - (2\lambda + \mu)\,\kappa_P^{(ab)}\,f_{3\pi} \,\sqrt{-\frac{\uh}{\sh}}
    \left(\frac{e_a}{\sh} + \frac{\sh}{\uh}\,\frac{e_b}{\uh} \right) \nn\\
    && \times\; \int_0^1\,\frac{d\tau}{\taub^2}\, \int_0^{\taub}\,\frac{d\tau_g}{(\taub-\tau_g)}\,
                 \phi_{3\pi}(\tau,\taub-\tau_g,\tau_g)\,.
\label{eq:DVMP-CG-CF}
\ea   
In the DVMP limit only the helicity non-flip amplitude ${\cal H}_{0-,++}^{P,tw3}$ is left over 
while ${\cal H}_{0-,-+}^{P,tw3}$ vanishes $\propto \th$ as a
consequence of angular momentum conservation. 
Thus, a full twist-3 DVMP analysis modifies the WW approximation
employed in \ci{GK5,GK6} by a change of the \da{} $\phi_{\pi p}$ generated through the 3-body \da{} via the
equation of motion \req{eq:phieqmotion} and by the additional 3-body contributions.
From the properties of the 3-body \da, $\phi_{3\pi}$, ( see Eq.\ \req{eq:3-particle-da}) it is clear that
the 3-body contributions~\footnote
   {Note that the 2-body twist-3  contribution proportional to $\phi_{\pi 2}^{EOM}$
and contributing to ${\cal H}^{P,C_F,\phi_{3 \pi}}$ 
vanishes for $\th\to 0$.} 
given in \req{eq:DVMP-CG-CF}
do not have end-point singularities. On the other hand, the $\phi_{\pi p}$ contribution possesses
an end-point singularity since this \da{} does not vanish at the end points as we already remarked. This
singularity has been regularized in \ci{GK5,GK6} by keeping the quark transverse momentum in the propagators.
More details on the twist-3 contribution in the DVMP limit will be given in a forthcoming paper \ci{DVMP}.

%%%%%%%%%%%%%%%%%%%%%%%%%%%%%%%%%%%%%%%%%%%%%%%%%%%%%%%%%%%%%%%%%%%%%%%%%%%%%%%%%%%%%%
\section{\da s, form factors and parameters}
\label{sec:DA}
%%%%%%%%%%%%%%%%%%%%%%%%%%%%%%%%%%%%%%%%%%%%%%%%%%%%%%%%%%%%%%%%%%%%%%%%%%%%%%%%%%%%%%

For the soft physics input to our calculation of pion electroproduction we use the same \da s
and form factors as for $\pi^0$ photoproduction \ci{KPK18}.
Thus, for the twist-2 pion \da{} we use the truncated Gegenbauer expansion
\be
\phiDA_\pi(\tau)\=6\tau\taub\,\big[1 + a_{2}\;C_2^{3/2}(2\tau -1)\big]
\label{eq:pionDA}
\ee
with the recent lattice QCD result on the second Gegenbauer coefficient \ci{braun15}
\be
a_{2}(\mu_0)\=0.1364 \pm 0.0213
\ee
at the initial scale $\mu_0=2$ GeV. The coefficient $a_2$ as all other expansion coefficients depend
on the factorization scale, $\mu_F$, which is defined by \req{eq:muRmuF}. 
The corresponding anomalous dimensions are quoted in App.\ A.

For the 3-body twist-3 \da we follow \ci{braun90} and use the  truncated conformal expansion   
\ba
\phiDA_{3\pi}(\tau_a,\tau_b,\tau_g)&=& 360\tau_a\tau_b\tau_g^2\Big[1 
                  + \omega_{1,0}\,\frac12(7\tau_g-3)\nn\\
             &+& \omega_{2,0}\,(2-4\tau_a\tau_b-8\tau_g+8\tau_g^2)   \nn\\
             &+& \omega_{1,1}\,(3\tau_a\tau_b-2\tau_g+3\tau_g^2)\Big]\,.\,
\label{eq:3-particle-da}
\ea
with
\be
\omega_{10}(\mu_0)\= -2.55\,, \qquad \omega_{20}(\mu_0)\=8.0\,, \qquad \omega_{11}(\mu_0)\=0.0\,.
\ee
The coefficients $\omega_{20}$ and $\omega_{11}$ mix under evolution, see App.\ A. The 3-body \da
 is normalized as 
\be
\int_0^1 d\tau \int_0^{\taub} d\tau_g \phi_{3\pi}(\tau,\taub-\tau_g,\tau_g) \= 1\,.
\ee
For the parameter $f_{3\pi}$ we take
\be
f_{3\pi}(\mu_0) \= 0.004\,\gev^2\,.
\ee
This parameter as well as the expansion coefficient $\omega_{10}$ have been derived from
QCD sum rules \ci{ball98}. According to \ci{ball98} the uncertainties of these parameters are large
of the order of $30\%$. The expansion coefficient $\omega_{20}$ has been adjusted by us \ci{KPK18}
to the CLAS data on $\pi^0$ photoproduction \ci{clas-pi0}.

Using \req{eq:3-particle-da}, we obtain the function $\phi_{\pi 2}^{EOM}$ from the
integral \req{eq:phi-EOM} 
\ba
\phi_{\pi 2}^{EOM}(\tau)&=& 120\,\frac{f_{3\pi}}{f_\pi\mu_\pi}\,\tau(1-\tau)^3 \Big[1 + \frac14\,\omega_{1,0}\, (1-7\tau) \nn\\
  &+& \frac25\,\omega_{2,0}\, (1-7\tau + 11\tau^2)
  - \frac1{10}\,\omega_{1,1}\, (1-7\tau + 6\tau^2) \big]\,.  
\ea

The equations of motion \req{eq:phieqmotion} can be suitably combined and solved for 
$\phi_{\pi p}$ and $\phi_{\pi\sigma}$ \ci{KPK18}:
\ba
\phi_{\pi\sigma}(\tau)&=& 6\tau\taub \left(\int d\tau \frac{\taub \phi^{EOM}_{\pi 2}(\taub)-\tau \phi_{\pi 2}^{EOM}(\tau)}
                                {2\tau^2\taub^2} + C\right)\,, \nn\\
\phi_{\pi p}(\tau)&=&\frac1{6\tau\taub} \phi_{\pi \sigma}(\tau) + \frac1{2\tau} \phi_{\pi 2}^{EOM}(\taub)
                               + \frac1{2\taub}\phi_{\pi 2}^{EOM}(\tau)\,.   
\label{eq:solution}
\ea
The constant of integration, $C$, is fixed from the constraint
\be
\int_0^1 d\tau \phi_{\pi p}(\tau)\=1\,.
\ee
Thus, for a given 3-body \da, $\phi_{3\pi}$, the 2-body twist-3 \da s are uniquely fixed. Since we started
from a truncated expansion of $\phi_{3\pi}$ we arrive at truncated Gegenbauer expansions of the 2-body
twist-3 \da s \cite{KPK18} (the $C_n^m$ denote the Gegenbauer polynomials):
\ba
\phi_{\pi p}(\tau)&=& 1+\frac17\frac{f_{3\pi}}{f_\pi\mu_\pi}\Big( 7\,\omega_{1,0}- 2\,\omega_{2,0}-\omega_{1,1}\Big) \nn\\
        && \hspace*{0.1\tw} \times\; \Big( 10\, C_2^{1/2}(2\tau-1) - 3\,C_4^{1/2}(2\tau-1)\Big)\,.
\label{eq:2-particle-das}
\ea
The \da{} $\phi_{\pi\sigma}$ is not needed by us in this work explicitly. Thus, we refrain from quoting it here.
It can be found in \ci{KPK18}.
If $\phi_{3\pi}=0$ Eq.\ \req{eq:solution} reduces to the well-known WW approximation of the \da s
\be
\phi_{\pi p}^{WW}\= 1\,, \qquad \phi_{\pi \sigma}^{WW}\=6\tau\taub\,.
\ee

Let us now turn to the discussion of the form factors, $F_i^P(t) (= R_i^P(t), S_i^P(t))$ which encode the
soft physics content of the nucleon matrix element. The form factors are given by process specific
combinations of the corresponding flavor form factors, $F_i^a$, which are defined as $1/x$-moments of
zero-skewness GPDs, $K_i^a$,
\be
F_i^a(t) \= \int_0^1 \frac{dx}{x}\, K_i^{a}(x,t) \,.
\label{eq:flavor-FF}
\ee
Here, $x=(k_j+k'_j)^+/(p+p')^+$ is the average momentum fraction the two active quarks carry and
$a$ is a valence quark. The restriction to valence quarks is an assumption for charged pions which will be
justified at the end of this section. For $\pi^0$-production this restriction
is a consequence of charge-conjugation parity. 

The usual GPDs, $K_i$, parameterize the soft proton-proton matrix elements. However, for electroproduction
of charged pions  proton-neutron transition matrix elements appear. With the help of SU(3) flavor symmetry
these transition GPDs and, hence, the corresponding form factors, can be related to the diagonal proton-proton
ones \ci{frankfurt}:
\ba
\gamma^*p\to \pi^+n:&& F_i^{\pi^+}(t)\=F_{i p\to n}(t)\=F_i^u(t)-F_i^d(t)\,, \nn\\
\gamma^*n\to \pi^-p:&& F_i^{\pi^-}(t)\=F_{i n\to p}(t)\=F_i^u(t)-F_i^d(t)\,. 
\label{eq:pion-FF}
\ea
 For $\pi^0$ production there are two subprocesses $\gamma^*u\to \pi^0 u$
and $\gamma^*d\to \pi^0 d$. In both cases $e_a=e_b$. It is therefore convenient to pull out
the charges from the subprocess amplitudes for this process and to absorb them into the form factors
together with the corresponding flavor weight factors \req{eq:flavor-weights}.
Thus, the form factors specific to $\pi^0$ production read
\be
F_i^{\pi^0}(t) \= \frac1{\sqrt{2}} \big[e_u F_i^u(t) - e_d F_i^d(t) \big]\,.
\ee
Consequently, the subprocess amplitudes do not depend anymore on the flavors in the case of $\pi^0$
production.

In \ci{DK13} the GPDs $H$ and $E$ for valence quarks have been extracted from the data
on the magnetic and electric form factors of the nucleon exploiting the sum rules for the form factors
with the help of a parameterization of the zero skewness GPDs
\be
K_i^a\=k_i^a(x) \exp{[tf_i^a(x)]}\,.
\label{eq:GPD-ansatz}
\ee
In \ci{DFJK4,DK13} it is advocated for the following parameterization of the profile function~\footnote
  {This ansatz is now supported by light-front holographic QCD \ci{teramond}. A similar parameterization
   has been proposed in \ci{moutarde}.}
\be
f_i^a(x)\=\big(B_i^a - \alpha_i'{}^a\ln{x}\big)(1-x)^3 + A_i^a x(1-x)^2
\,,
\label{eq:profile}
\ee
with the parameters $A_i$, $B_i$ and $\alpha_i$ fitted to the data of the nucleon's electromagnetic form factors.
The forward limit of the GPD $H^a$ is given by the flavor-a parton density, $q^a(x)$ which is taken from
\ci{abm11}. Since the forward limit of $E^a$ is not accessible in deep-inelastic scattering it is,
therefore, to be determined in the form factor analysis, too. The most prominent feature of the GPDs,
parameterized as in \req{eq:GPD-ansatz} and \req{eq:profile}, is the strong $x - t$ correlation
\ci{DFJK4,DK13}: The GPDs at small $x$ control the behavior of their associated flavor form factors at 
small $-t$ whereas large $x$ determine their large $-t$ behavior. The flavor
form factors $R_V^a$ and $R_T^a$ are evaluated from the GPDs \req{eq:GPD-ansatz}, \req{eq:profile}. These
flavor form factors have also been used in wide-angle Compton scattering \ci{DK13}.

For the form factor, $R_A$, being related to the GPD $\widetilde{H}$, we use example \#1 discussed
in  \ci{kroll17}. This example is extracted from the data on the axial form factor of the nucleon and on
the helicity correlations, $A_{LL}$ and $K_{LL}$, measured in wide-angle Compton scattering \ci{hamilton,fanelli}.

For the transversity GPDs $H_T$ and $\bar{E}_T$ the parameterization \req{eq:GPD-ansatz}, \req{eq:profile} is
also employed and the parameters, $B_i^a$ and $\alpha'_i{}^{a}$, fixed from the low $-t$ data on
deeply-virtual pion electroproduction \ci{GK5,GK6}. The  values of these parameters can be found in \ci{KPK18}.
For wide-angle photo- and electroproduction the large $-t$ behavior, i.e.\ the second term in the profile function,
is also required. As in \ci{KPK18} we use for the relevant parameter $A_i^a$ the value $0.5\,\gev^2$ for all
transversity GPDs. With this choice a good fit to the CLAS data \ci{clas-pi0} on $\pi^0$ photoproduction has been
obtained.
At present there is no information available on the GPD ${\widetilde H}_T$ and its associated from 
factor $S_S$. In order to have an at least rough estimate of its importance we assume, with regard to the
definition of the GPD $\bar{E}_T$ 
\be
\bar{E}_T\=2\widetilde{H}_T + E_T\,,
\ee
that $S_S^a=\bar{S}^a_T/2$. This assumption is equivalent to the neglect of $E_T$.

\begin{table*}[t]
\renewcommand{\arraystretch}{1.4} 
\caption{The powers $d_i$ for the form factors contributing
  to the wide-angle photo- and electroproduction of pions. The table is taken
  from ref.\ \ci{KPK18}.} 
\begin{center}
\begin{tabular}{| c || c  c  c  c  c|}
     & $R^a_V$ & $R^a_A$ &  $R^a_T$ &  $S^a_T$ & $\bar{S}^a_T$ \\[0.2em]
\hline   
 $u$ & 2.25  & 2.22 &  2.83    & 2.5 & 2.5   \\[0.2em]
 $d$ & 3.0   & 2.61 &  3.12    & 3.5 & 3.0    \\[0.2em]
\hline
\end{tabular}
\end{center}
\label{tab:3}
\renewcommand{\arraystretch}{1.0}   
\end{table*} 

Last not least we want to discuss a property of the GPDs \req{eq:GPD-ansatz}, \req{eq:profile}
which is of particular significance for exclusive wide-angle processes.
With the help of the saddle point method \ci{DFJK4} one can show  that moments of these GPDs, 
fall as a  power of $t$ at large $-t$:
\be
    F^a_i \sim 1/(-t)^{d^{a}_i}
\label{eq:ff-power}
\ee
where
\be
d^a_i=(1+\beta^a_i)/2\,.
\ee
and $\beta^a_i$ is the power of $1-x$ with which the forward limit of the GPD $K_i^a$ vanishes for $x \to 1$.
Eq.\ \req{eq:ff-power} is a generalization of the famous Drell-Yan relation \ci{DY}.
The phenomenological values of the powers $d_i^a$ are listed in Tab.\ \ref{tab:3}. Note the differences
between the powers of $u$ and $d$-quarks. This implies that at very large $-t$ only $u$-quark flavor form factors
contribute. In case of the helicity non-flip GPDs the powers $\beta_i^u$ are slightly larger than expected
according to perturbative QCD arguments \ci{brodsky94,yuan03}.
We however stress that, in practice, the powers, $\beta^a_i$, are fixed in a region of $x$ less than about
0.8 for the helicity non-flip GPDs,  and even a smaller $x$-region for the transversity ones.
For larger $x$ there is no experimental information on the forward limits available at present. Therefore, the 
powers $\beta^a_i$ are to be considered as effective powers which are likely subject to 
changes as soon as data at larger $x$ become available. These powers affect the energy dependence of the
cross sections. For photoproduction the energy dependence of the cross section at fixed $\cos{\theta}$
can readily be read off from \req{eq:amplitudes} and the subprocess amplitudes discussed in Sect.\
\ref{sec:subprocess}. One finds the familiar $1/s^7$ scaling behavior of the cross section at fixed
$\cos{\theta}$ for the twist-2 contribution
and $1/s^8$ for the twist-3 one provided the form factors, including the prefactors of $\sqrt{-t}$ and $t$
appearing in \req{eq:amplitudes}, drop as $1/t^2$. Deviations from that behavior change the scaling behavior as
do the logs from $\als$ and the evolution of the \da{} parameters. As can be seen from Tab.\ \ref{tab:3}
our form factors increase the power of $s$ with which the cross section falls. In the energy range we explore
the effective scaling is $1/s^9$.

As is well-known the sea-quark densities fall faster to zero for $x\to 1$, typically as $\sim (1-x)^7$ than the
valence quark densities, see for instance \ci{abm11}. This is also expected from perturbative QCD arguments
\ci{brodsky94}. The forward limits of the other GPDs may fall even more rapidly than that of $H^{\rm sea}$.
According to \req{eq:ff-power} the sea-quark form factor falls as
\be
R_V^{\rm sea}(t) \sim 1/(-t)^4\,.
\ee
The other sea-quark form factors decrease like $R_V^{\rm sea}$ or even faster. Thus, we conclude that sea-quark
contributions to wide-angle electroproduction are strongly suppressed and therefore neglected by us.

%%%%%%%%%%%%%%%%%%%%%%%%%%%%%%%%%%%%%%%%%%%%%%%%%%%%%%%%%%%%%%%%%%
\section{Photoproduction}
\label{sec:photoproduction}
%%%%%%%%%%%%%%%%%%%%%%%%%%%%%%%%%%%%%%%%%%%%%%%%%%%%%%%%%%%%%%%%%%

\begin{figure}[p]
\centering
  \includegraphics[width=0.48\tw]{fig-dsdt-1.epsi}
  \includegraphics[width=0.48\tw]{fig-dsdt-piminus-11.epsi}
  \caption{\label{fig:dsdt-pi0} Left: The cross section for $\pi^0$ photoproduction versus the cosine of the c.m.s.
    scattering angle, $\theta$, at $s=11.06\,\gev^2$. The solid (dotted) curve represents the full (twist 2) result
    using the same parameters as in \ci{KPK18}. 
    The red dashed curve is obtained with the same parameters as in \ci{KPK18} but with the amplitudes taken at the 
    fixed scale $\mu_R=\mu_F=1\,\gev$, while for the blue dashed curve we additionally change $\omega_{20}=10.3$.
    Data taken from \ci{anderson76} (full circles) at $s=10.3\,\gev^2$ and from CLAS \ci{clas-pi0} (open circles)
    at $s=11.06\,\gev^2$. The cross sections are scaled by $s^7$  and the theoretical results are only shown for
    $-t$ and $-u$ larger than $2.5\,\gev^2$. Right: Results for the $\pi^-$ photoproduction cross section at
    $s=11.3\,\gev^2$. Data, shown as open triangles, are from \ci{zhu05}.  For other notations it is referred to
    the figure on the left hand side.}
    \includegraphics[width=0.48\tw]{fig-dsdt-piplus-10.epsi}
  \includegraphics[width=0.48\tw]{fig-dsdt-piplus-15.epsi}
  \caption{\label{fig:dsdt-pi+} Results for the  $\pi^+$ photoproduction cross sections vs. $\cos{\theta}$, at
    $s=10.3\,\gev^2$ (left) and $15\,\gev^2$ (right). The full circles are the data from \ci{anderson76} at
    $s=10.3$ and $15\,\gev^2$; the open triangles are from 
   \ci{zhu05} at $s=11.3\,\gev^2$. At $s=15\,\gev^2$ results are shown for $-t$ and $-u$ larger than $4\,\gev^2$.
   For other notations it is referred to Fig.\ \ref{fig:dsdt-pi0}.  }
\end{figure}

The photoproduction cross section
\be
\frac{d\sigma^P}{dt} \= \frac1{32\pi (s-m^2)^2}\,\sum_{\nu'\mu} |{\cal M}^P_{0\nu',\mu+}|^2 
\label{eq:cross-section}
\ee
is evaluated from \req{eq:amplitudes} using the subprocess amplitudes \req{eq:Htw2Tphoto1}
and \req{eq:Htw3Tphoto} as well as the \da s and form factors described in Sec.\ \ref{sec:DA}.
The resulting cross sections for the various pion channels, scaled by $s^7$ in order to take away
most of the energy dependence,
are  displayed in Figs.\ \ref{fig:dsdt-pi0} and \ref{fig:dsdt-pi+} as solid lines and compared
to experiment \ci{clas-pi0,anderson76,zhu05}. Of course, there is agreement with the CLAS data
\ci{clas-pi0} since the expansion coefficient $\omega_{20}$ is fitted to these data. On the other
hand, there is substantial disagreement with the SLAC data \ci{anderson76}. In particular the
SLAC $\pi^0$ data are about an order of magnitude larger than the new CLAS data in the vicinity
of $90$ degrees. With regard to the prerequisite of the handbag approach that the  Mandelstam
variables should be much larger than the hadronic scale, $\Lambda^2$, we only show results for
$-t$ and $-u$ larger than $2.5\,\gev^2$. This is a compromise between the requirement of the
handbag approach, on the one side, and having available a not too small range of $\cos{\theta}$
for the experimentally accessible values of $s$ on the other side.

The corresponding results for the $\pi^\pm$ cross sections are shown in Figs.\ \ref{fig:dsdt-pi0}
and \ref{fig:dsdt-pi+} and compared to the SLAC \ci{anderson76} and Jefferson lab Hall A data \ci{zhu05}.
For positive values of $\cos{\theta}$ theory and experiment are close to each other  but near $90$
degrees and in the backward hemisphere our results are well below the data \ci{anderson76,zhu05}
(by a factor of 2 to 3). In contrast to $\pi^0$ photoproduction the twist-2 contribution to the
$\pi^\pm$ cross section is substantial in the forward hemisphere, see Figs.\ \ref{fig:dsdt-pi0} and
\ref{fig:dsdt-pi+}. This can be understood from properties of the subprocess amplitude \req{eq:Htw2Tphoto1}:
\be
\frac{{\cal H}^{\pi^\pm}}{{\cal H}^{\pi^0}} \sim \frac{e_a \uh +e_b\sh}{\sh+\uh}
\ee
This ratio is large for small $-\th=-\sh-\uh$. With rising $-t$, i.e. decreasing $\cos{\theta}$, the
twist-3 contribution quickly takes the lead. The twist-2 - twist-3
interference is however noticeable
in the entire wide-angle region even for $\pi^0$ photoproduction. At $\cos{\theta}\simeq -0.4$ the
interference term still amounts to about $10\%$, positive for $\pi^+$ production and negative for
the case of $\pi^-$. With increasing $s$ the twist-2 contribution becomes more important
(see Fig.\ \ref{fig:dsdt-pi+}) since, as is evident from Tab.\ \ref{tab:power-behavior-s}, the twist-3
subprocess amplitude is suppressed by an extra factor $1/\sqrt{\sh}$.

The $\pi^-$ photoproduction cross section is larger than the $\pi^+$ one by a factor 2-3. The reason for this
fact are the quark-charge factors in \req{eq:Htw2Tphoto1} and \req{eq:Htw3Tphoto} which favor the $\pi^-$
channel ($n=1,2$, see \req{eq:Htw3Tphoto})
\be
\frac{{\cal H}^{\pi^-}}{{\cal H}^{\pi^+}} \sim \frac{e_d\uh^n+e_u\sh^n}{e_u\uh^n+e_d\sh^n}\,.
\label{eq:P-M-ratio}
\ee
The absolute value of this ratio is larger than 1. 

The discrepancy between theory and experiment is larger at $s=15\,\gev^2$ than at $10.3\,\gev^2$. 
It seems that the predicted energy dependence is too strong. It also seems that the ratio of the $\pi^-$
and $\pi^+$ cross sections at $90^\circ$ is too large as compared to the Hall A data \ci{zhu05}.
Some fine tuning of the soft-physics input is perhaps necessary. We however hesitate to do so since
the SLAC data are very old and a remeasurement of wide-angle $\pi^\pm$  photoproduction seems to be advisable.

At the end of Sect.\ \ref{sec:DA} we already discussed the energy dependence of the handbag approach.
As we mentioned in Sec.\ \ref{sec:DA}, in the range of $s$ we are interested in, our cross section effectively
behaves $\propto s^{-9}$ in the region of twist-3 dominance. This is perhaps somewhat too strong.
Since our form factors represent $1/x$-moments of GPDs they evolve with the scale in principle. 
Because of the strong $x-t$ correlation the form factors at large $-t$ are under control of a narrow
region of large $x$. With increasing $-t$ the affected region approaches 1 and becomes narrower. Therefore,
our form factors approximately become equal to the scale-independent lowest moments of the GPDs concerned.
Thus, as it is argued in \ci{DFJK4}, the $1/x$-factors in the form factors can be viewed as a 
phenomenological estimate of effects beyond the strict $\Lambda/\sqrt{-t}$ expansion (see
Sect.\ \ref{sec:handbag}). One may likewise argue that the disregard of the scale-dependence of the
form factors also requires the neglect of the evolution of the \da s for consistency. Thus,
as already discussed in \ci{KPK18}, we also evaluate the photoproduction cross sections at the fixed scale
of $\mu_R=\mu_F=1\,\gev$ as an alternative and fit the coefficient $\omega_{20}$ to the CLAS and SLAC data.
This procedure hardly alters the size and shape of the cross sections but reduces their effective energy
dependence to about $s^{-8}$ in regions of twist-3 dominance. 
The results we obtain
with the fixed scale are shown as dashed lines in Figs.\ \ref{fig:dsdt-pi0}:
red for the usual parameters taken form \ci{KPK18} and blue for $\omega_{20}=10.3$.
They agree fairly well with the data and one notes that freezing the scale has bigger effect than changing
$\omega_{20}$.

\clearpage

%%%%%%%%%%%%%%%%%%%%%%%%%%%%%%%%%%%%%%%%%%%%%%%%%%%%%%%%%%%%%%%%%%%%%%%%%%%%%%%
\section{Electroproduction}
\label{sec:electroproduction}
%%%%%%%%%%%%%%%%%%%%%%%%%%%%%%%%%%%%%%%%%%%%%%%%%%%%%%%%%%%%%%%%%%%%%%%%%%%%%%%%

\begin{figure}[p]
\centering
\includegraphics[width=0.47\tw]{fig-dsdtL-T-15.epsi} \hspace*{0.03\tw}
\includegraphics[width=0.47\tw]{fig-dsdtT-15.epsi}
\caption{\label{fig:partial-cs-15-1} Left: Predictions for the ratio of the longitudinal and transverse cross sections
  for $\pi^+$ electroproduction versus the cosine of the c.m.s. scattering angle at $s=15\,\gev^2$ for a set
  of $Q^2$ values. Parameters as in \ci{KPK18}. The predictions are only shown for $-\th$ and $-\uh$ larger
  than $4\,\gev^2$. Right: As the figure on the left hand side but for the ratio of the transverse and the
  photoproduction cross sections.} 
{~}\\
\includegraphics[width=0.47\tw]{fig-dsdtLT-T-15.epsi} \hspace*{0.03\tw}
\includegraphics[width=0.47\tw]{fig-dsdtTT-T-15.epsi}
\caption{\label{fig:partial-cs-15-2} Predictions for the longitudinal-transverse (left) and transverse-transverse
  (right) interference cross sections of $\pi^+$ electroproduction divided by the transverse one  versus the
  cosine of the c.m.s. scattering angle at $s=15\,\gev^2$ for a set of photon virtualities.} 
\end{figure}

%\clearpage

\begin{figure}[t]
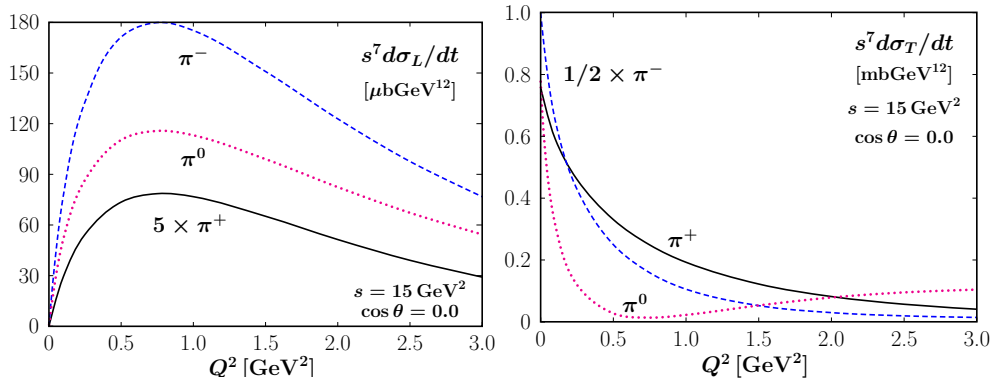

\centering
\includegraphics[width=0.47\tw]{fig-dsdt-L-15.epsi}
\includegraphics[width=0.47\tw]{fig-dsdt-T-15.epsi}
\caption{\label{fig:partial-cs-15-3} Predictions for the longitudinal (left) and the transverse (right)
  cross sections of $\pi^+$ (solid), $\pi^-$ (dashed) and $\pi^0$ (dotted line) electroproduction versus $Q^2$
  at $s=15\,\gev^2$ and $\cos{\theta}=0$. Parameters as in \ci{KPK18}.}
\end{figure}

\begin{figure}[p]
\centering
\includegraphics[width=0.47\tw]{fig-dsdtL-T-10.epsi} \hspace*{0.03\tw}
\includegraphics[width=0.47\tw]{fig-dsdtT-T-10.epsi}
\caption{\label{fig:partial-cs-10-1} Predictions for the ratio of the longitudinal and transverse cross
  sections (left) and the transverse cross sections divided by the photoproduction one (right)
  of pion electroproduction vs. $\cos{\theta}$ at $s=10.3\,\gev^2$ and $Q^2=2.0\,\gev^2$. 
  Parameters as in \ci{KPK18}. The predictions are only shown for $-\th$ and
  $-\uh$ larger than $2.5\,\gev^2$.}
%\end{figure}  
%\begin{figure}[t]
%\centering
{~}\\
\includegraphics[width=0.47\tw]{fig-dsdtLT-T-10.epsi} \hspace*{0.03\tw}
\includegraphics[width=0.48\tw]{fig-dsdtTT-T-10.epsi}
\caption{\label{fig:partial-cs-10-2} Predictions for the longitudinal-transverse (left) and transverse-transverse
  (right) interference cross sections of pion electroproduction vs. $\cos{\theta}$ at at $s=10.3\,\gev^2$ and
  $Q^2=2.0\,\gev^2$. The interference cross sections are divided by the corresponding transverse cross section.}
\end{figure}

%\clearpage
As is well known, for pion electroproduction there are four partial cross sections in contrast to just
Eq.\ \req{eq:cross-section} for photoproduction.
For the ease of access we repeat the definitions of the partial cross sections in terms of helicity amplitudes 
\ba
\frac{d\sigma_L}{dt}&=& 2\varrho_{ps} \left[ |{\cal M}_{0+,0+}|^2 + |{\cal M}_{0-,0+}|^2\right]\,, \nn\\
\frac{d\sigma_T}{dt}&=& \varrho_{ps} \sum_\mu\left[ |{\cal M}_{0+,\mu +}|^2 + |{\cal M}_{0-,\mu +}|^2\right]\,,\nn\\
\frac{d\sigma_{LT}}{dt}&=& -\sqrt{2} \varrho_{ps} {\rm Re}\,\sum_\mu\,\mu \left[{\cal M}^*_{0-,0+}{\cal M}_{0-,\mu +}
+ {\cal M}^*_{0+,0+}{\cal M}_{0+,\mu +} \right]\,, \nn\\
\frac{d\sigma_{TT}}{dt}&=& - 2\varrho_{ps} {\rm Re}\,\left[{\cal M}^*_{0-,++}{\cal M}_{0-,- +}  
+  {\cal M}^*_{0+,++}{\cal M}_{0+,- +} \right]\,,
\label{eq:partial-cs}
\ea
 where the phase space factor is given by
 \be
 \varrho_{ps}\=\Big[32\pi(s-m^2)\sqrt{\Lambda_M(s,-Q^2,m^2)}\Big]^{-1}\,,
   \ee
   where $\Lambda_M$ is the Mandelstam function~\footnote{
         In DVMP $s$ is usually denoted by $W^2$}.\\
   The partial cross sections sum up to the unpolarized $ep\to e\pi N$ cross section:
   \ba
   \frac{d^4\sigma}{dsdQ^2dtd\varphi}&=& \frac{\ale (s-m^2)}{16\pi^2E_L^2 m^2 Q^2 (1-\veps)}\, \left(
   \frac{d\sigma_T}{dt} + \veps \frac{d\sigma_L}{dt} \right. \nn\\
   & + & \left. \veps \cos{(2\varphi)}\, \frac{d\sigma_{TT}}{dt}
   + \sqrt{2\veps (1+\varepsilon)}\, \cos{\varphi}\, \frac{d\sigma_{LT}}{dt}\,\right)
   \ea
   Here, $\varphi$ is the azimuthal angle between the lepton and the hadron plane and $E_L$ is
   the energy of the lepton beam. The ratio of the longitudinal and transversal photon flux is denoted by $\veps$.

The twist-2 and twist-3 subprocess amplitudes for pion electroproduction
are presented in Sec. \ref{sec:subprocess}.  From these subprocess amplitudes in combination with
the form factors described in Sect.\ \ref{sec:DA}, we evaluate the amplitudes \req{eq:amplitudes}, 
\req{eq:long-amplitudes} and subsequently the partial cross sections \req{eq:partial-cs}.
In the light of the discussion in the preceding section we will mostly show ratios of cross sections for
pion electroproduction. Most of the energy dependence and of the normalization uncertainties cancel in the ratios.
Therefore, we only show results evaluated from the same \da s and flavor form factors as in \ci{KPK18}.

First, we compare the partial cross sections for, say, $\pi^+$ production for different photon
virtualities. In order to have at disposal a rather large range of $Q^2$ we need large $-t$ and $-u$ because of
the requirement \req{eq:condition-1} and consequently large $s$. Therefore, we choose the not unrealistically
large value of $15\,\gev^2$ for $s$ and show the partial cross sections only for $Q^2=1,\,2$ and $3\,\gev^2$
and $-t,-u \geq 4\,\gev^2$. In Fig.\ \ref{fig:partial-cs-15-1} we present the ratio of the longitudinal and the
transverse cross section as well as the ratio of the transverse and the photoproduction cross section for
$Q^2=1,\,2$ and $3\,\gev^2$. The interference cross sections, divided by the transverse one, are displayed in
Fig.\ \ref{fig:partial-cs-15-2}.
The ratios reveal a mild $\cos{\theta}$ and $Q^2$ -dependence. The latter one is getting only
   somewhat stronger in the backward hemisphere. 

In Fig.\ \ref{fig:partial-cs-15-3}
   we show the separate longitudinal and transverse cross sections  
   versus $Q^2$ at $\cos{\theta}=0$. Starting at zero in the photoproduction limit the longitudinal cross section
   increases with rising $Q^2$ up to a maximum at about $1.0\,\gev^2$ while the transverse cross sections for
   charge pions are continuously decreasing. The $\pi^0$ cross section has a mild minimum at about $Q^2=0.7\,\gev^2$.
The magnitudes of the longitudinal cross
sections differ markedly. The $\pi^+$ cross section is very small compared to the other ones. As for
photoproduction, see Eq.\ \req{eq:P-M-ratio}, the quark charges favor the $\pi^-$ production over the $\pi^+$ one.

In  Figs.\ \ref{fig:partial-cs-10-1} and  \ref{fig:partial-cs-10-2} the partial cross sections for the three pion
channels are shown at $s=10.3\,\gev^2$ and at a fixed $Q^2$ of $2\,\gev^2$. As in Figs.\ \ref{fig:partial-cs-15-1}
and \ref{fig:partial-cs-15-2} the partial cross sections are divided by the transverse one except of the
transverse cross section itself which is divided by the corresponding photoproduction cross section. The cross
sections for the various pions differ markedly from each other. Particularly noteworthy are the maxima in the
partial cross sections occurring near $90$ degrees for $\pi^-$ and at about $115$ degrees for $\pi^0$ production.
The sharp peak of the ratio of $d\sigma_L$ and $d\sigma_T$, especially for the case of $\pi^0$ production, is
generated by a conspiracy of minima at slightly different positions and dissimilar depths of these cross sections.
Comparison of the $\pi^+$-curves in Figs.\ \ref{fig:partial-cs-10-1} and  \ref{fig:partial-cs-10-2}
with the $Q^2=2\,\gev^2$
curves in Figs.\ \ref{fig:partial-cs-15-1} and \ref{fig:partial-cs-15-2} 
gives an impression of the energy dependence of pion electroproduction.

Data on the partial cross section of pion electroproduction will allow for an extraction of detailed information
on the large $-t$-behavior of the transversity GPDs. Thus, as is evident from \req{eq:long-amplitudes}, the
longitudinal cross section is only dependent upon the form factors $(R_A^P)^2$ and $(S_T^P)^2$; there is no
interference between the twist-2 and twist-3 contributions. As a little calculation reveals, the
longitudinal-transverse interference cross section has the same structure:
\ba
\frac{d\sigma^P_{LT}}{dt}&=& -\sqrt{2} e_0^2 \varrho_{ps} \Big[ {\cal H}^P_{0+,0+} \big({\cal H}^P_{0+,++}
                         - {\cal H}^P_{0+,-+}\big) (R_A^P)^2 \nn\\
         && \hspace*{0.11\tw} +\, {\cal H}^P_{0-,0+} \big({\cal H}^P_{0-,++} - {\cal H}^P_{0-,-+}\big) (S_T^P)^2 \Big]
\ea
Given that the axial form factor, $R_A^P$, is not unknown at large $-t$, from data on the longitudinal and the
longitudinal-transverse cross sections we may extract information on $S_T^P$ and thus on $H_T$
from data on the longitudinal and the
longitudinal-transverse interference cross sections. The transverse as well as the transverse-transverse interference
cross sections depend on all six form factors and, hence, on the form factor $S_S^P$ too. This form factor represents
the $1/x$-moment of the completely unknown transversity GPD $\widetilde{H}_T$. In DVMP this GPD is strongly suppressed
since it comes together with the factor $t/(4m^2)$. Numerical examination laid open that $d\sigma_{TT}^P$ is very
sensitive to $S_S^P$ in the regions of twist-3 dominance. Thus, it seems that a measurement of $d\sigma_{TT}^P$
may provide information on $\widetilde{H}_T$ at least at large $-t$.

\clearpage

%%%%%%%%%%%%%%%%%%%%%%%%%%%%%%%%%%%%%%%%%%%%%%%%%%%%%%%%%%%%%%%%%%%%%%%%%%%%
\section{Spin effects}
\label{sec:spin}
%%%%%%%%%%%%%%%%%%%%%%%%%%%%%%%%%%%%%%%%%%%%%%%%%%%%%%%%%%%%%%%%%%%%%%%%%%

The derivation of the photo- and electroproduction  amplitudes within the handbag approach naturally
requires the use of the light-cone helicity basis. However, for comparison with experimental
results on spin-dependent observables, the use of ordinary photon-nucleon c.m.s. helicity amplitudes
is more convenient. The ordinary helicity amplitudes, $\Phi_{0\nu',\mu\nu}$,
are obtained from the light-cone ones  \req{eq:amplitudes}, 
by the transform  (see  \ci{diehl01,pauli01,HKM})
\ba
\Phi^P_{0\nu',\mu\nu}&=& {\cal M}^P_{0\nu',\mu\nu}  \nn\\
                  &+& (-1)^{1/2+\nu'}\,\kappa'\,{\cal M}^P_{0-\nu',\mu\nu}
                     + (-1)^{1/2+\nu}\,\kappa\,{\cal M}^P_{0\nu',\mu-\nu} + {\cal O}(m^2/s)\,, \nn\\
\Phi_{0\nu',0\nu}&=& {\cal M}^P_{0\nu',0\nu}
                   + (-1)^{1/2+\nu'}\,\big(\kappa'+\kappa\big)\,{\cal M}^P_{0-\nu',0\nu} + {\cal O}(m^2/s)                   
\label{eq:helicity-transform}
\ea
where
\ba
\kappa&=&\frac{m}{s+Q^2}\,\sqrt{s+Q^2/2}\,\frac{\sin{\vartheta}}{1+\cos{\vartheta}}\,, \nn\\
\kappa'&=& -\kappa\Big(1 - \frac{Q^2}{4s} - \frac{Q^2}{s}\,\frac{\cos{\vartheta}}{1-\cos{\vartheta}}\Big)
\ea
For convenience the notation of the helicities is kept and the angle $\vartheta$ is defined in \req{eq:varphi}.
In the photoproduction limit $\kappa'$ becomes $-\kappa$ and
\be
\kappa \stackrel{Q^2\to 0}{\longrightarrow} \frac{m}{\sqrt{s}}\,\frac{\sqrt{-t}}{\sqrt{s}+\sqrt{-u}}
\label{eq:photo-hel-transform}
\ee
Obviously,
\ba
\sum_{\nu',\mu}|\Phi^P_{0\nu',\mu +}|^2&=&\sum_{\nu',\mu}|{\cal M}^P_{0\nu',\mu +}|^2\,, \nn\\
\sum_{\nu'}|\Phi^P_{0\nu',0 +}|^2&=&\sum_{\nu'}|{\cal M}^P_{0\nu',0 +}|^2\,.
\ea

%%%%%%%%%%%%%%%%%%%%%%%%%%%%%%%%%%%%%%%%%%%%%%%%%%%%%%%%%%%%%%%%%%%%%%%%%%%%%%%%%%%%%%%%%
\subsection{Helicity correlations in photoproduction}
\begin{figure}[t]
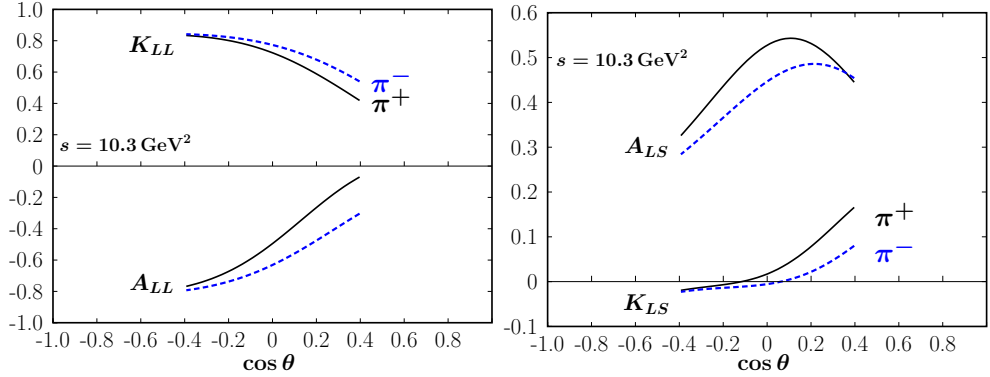

\begin{center}
\includegraphics[width=0.47\tw]{fig-ALL-KLL-pi-photo.epsi}
\includegraphics[width=0.47\tw]{fig-ALS-KLS-pi-photo.epsi}
\caption{\label{fig:spin-piP} Results for the helicity correlation parameters $A_{LL}, K_{LL}$ (left)
  and $A_{LS}, K_{LS}$ (right) for $\pi^+$ and $\pi^-$ photoproduction vs. $\cos{\theta}$ at $s=10.3\,\gev^2$.
Parameters as in \ci{KPK18}.}
\end{center}
\end{figure}

%%%%%%%%%%%%%%%%%%%%%%%%%%%%%%%%%%%%%%%%%%%%%%%%%%%%%%%%%%%%%%%%%%%%%%%%%%%%%%%%%%%%%%%%%
As for wide-angle Compton scattering the most interesting spin-dependent 
observables of pion photoproduction are the correlations of the helicities of the incoming photon and
that of either the incoming or the 
outgoing nucleon, $A_{LL}$, or $K_{LL}$, respectively. In terms of helicity amplitudes these
observables are defined by
\ba
A^P_{LL} &=& \frac{|\Phi^P_{0+,++}|^2 - |\Phi^P_{0+,-+}|^2 + |\Phi^P_{0-,++}|^2 - |\Phi^P_{0-,-+}|^2 }
                {\sum_{\nu',\mu}|\Phi^P_{0\nu',\mu +}|^2}\,, \nn\\
K^P_{LL} &=& \frac{|\Phi^P_{0+,++}|^2 - |\Phi^P_{0+,-+}|^2 - |\Phi^P_{0-,++}|^2 + |\Phi^P_{0-,-+}|^2 }
                {\sum_{\nu',\mu}|\Phi^P_{0\nu',\mu +}|^2}\,. 
\label{eq:correlations}
\ea
One can easily check that for the twist-3 contribution one has 
\be
A_{LL}^{P,tw3}\=-K_{LL}^{P,tw3}
\ee
while for twist 2 
\be
A_{LL}^{P,tw2}\=\phantom{-}K_{LL}^{P,tw2}
\ee
holds as is the case for wide-angle Compton scattering \ci{kroll17,HKM}. Thus, the experimental
observation of an approximate mirror symmetry between $A_{LL}^P$ and $K_{LL}^P$ signals the dominance
of twist-3 contributions to pion photoproduction. With regard on that feature  $A^P_{LL}$ and $K^P_{LL}$
play a similar important role for the discrimination between twist 2 and twist 3 in photoproduction of pions as 
the longitudinal and transverse cross sections in DVMP.
In Fig.\ \ref{fig:spin-piP} we show the helicity correlations for $\pi^\pm$ photoproduction at $s=10.3\,\gev^2$.
They are very similar for these cases.
In the backward hemisphere $A^P_{LL}$ and $K^P_{LL}$ are mirror symmetric which reflects
the twist-3 dominance there. On the other hand, in the forward hemisphere where the twist-2 contribution becomes
increasingly more significant $A_{LL}$ and $K_{LL}$ approach each other. The energy dependence of the helicity
correlation parameters is very  weak.
$\pi^0$ photoproduction has been discussed by us in \ci{KPK18} in great detail. The helicity correlations
for this channel reveal the approximate mirror symmetry in the forward hemisphere too since the twist-2 contribution
is also tiny in that region, see Fig.\ \ref{fig:dsdt-pi0}.

As one sees from Eq.\ \req{eq:Htw3Tphoto}  there is only one independent non-vanishing twist-3 subprocess amplitude,
namely ${\cal H}^{P,tw3}_{0-,-+}$ ( $=-{\cal H}^{P,tw3}_{0+,+-}$ by parity conservation). This amplitude therefore cancels
in \req{eq:correlations} and, to twist-3 accuracy,  the helicity correlations are solely expressed by the transversity
form factors. In particular the numerator reads up to corrections of order $\kappa$  
\be
A_{LL}^{P,tw3}\=-K_{LL}^{P,tw3} \propto - S_T^P\Big( S_T^P - \frac{t}{2m^2}\,S_S^P\Big)\,.
\ee

Another spin-dependent observable is the correlation between the helicity of the incoming photon and
the sideways polarization (i.e.\ the polarization perpendicular to the nucleon momentum but in the scattering plane)
of the incoming ($A^P_{LS}$) or outgoing ($K^P_{LS}$) nucleon
\ba
A_{LS}^P &=& 2\, \frac{{\rm Re}\, \Big[\Phi^{P*}_{0+,++}\Phi^P_{0-,-+} - \Phi^{P*}_{0+,-+} \Phi^P_{0-,++}\Big]}
                                  {\sum_{\nu',\mu}|\Phi^P_{0\nu',\mu +}|^2}\,,\nn\\
K_{LS}^P &=&2\,\frac{{\rm Re}\,\Big[ \Phi^{P*}_{0+,++}\Phi^P_{0-,++} - \Phi^{P*}_{0+,-+} \Phi^P_{0-,-+}\Big]}
                                  {\sum_{\nu',\mu}|\Phi^P_{0\nu',\mu +}|^2}\,.
\ea
Predictions for these observables are displayed in Fig.\ \ref{fig:spin-piP} as well. Both twist 2 and twist 3
contribute substantially to these observables. The order $m/\sqrt{s}$ mass corrections \req{eq:photo-hel-transform}
are also rather large for the $LS$ correlations. In addition $A_{LS}$ and $K_{LS}$ are subject to strong cancellations
among the contributions from various helicity amplitudes. In contrast to the helicity correlations, $A_{LL}$ and
$K_{LL}$, the mirror symmetry is therefore not to be seen for the $LS$ correlations.

%%%%%%%%%%%%%%%%%%%%%%%%%%%%%%%%%%%%%%%%%%%%%%%%%%%%%%%%%%%%%%%%%%%%%%%%%%%%%%%%%%%
\subsection{Helicity correlations in electroproduction}

\begin{figure}[t]
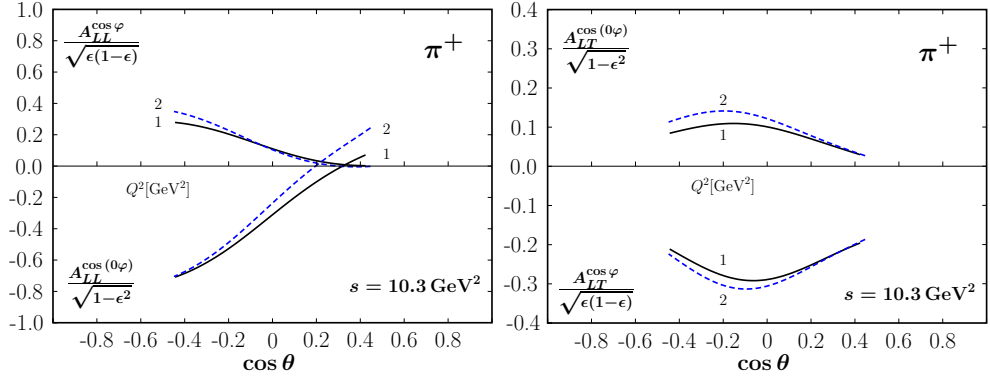

\begin{center}
\includegraphics[width=0.47\tw]{fig-ALL-pion-electro-Q2.epsi}
\includegraphics[width=0.47\tw]{fig-ALT-pion-electro-Q2.epsi}
\caption{\label{fig:spin-pi-electro} Results for the modulations of the helicity correlation $A_{LL}$ (left)
  and the observable $A_{LT}$ (right) for $\pi^+$ electroproduction vs $\cos{\theta}$ at $Q^2=1$(solid) and
  $2\,\gev^2$ (dashed lines) at $s=10.3\,\gev^2$.}
\end{center}
\end{figure}

The helicity correlation $A_{LL}$ can also be measured in pion electroproduction. In fact, the CLAS collaboration
has measured it in the deeply virtual region \ci{CLAS-ALL}. In electroproduction there are two modulations of
$A_{LL}$. Its $\cos{(0\varphi)}$ modulation, divided by $\sqrt{1-\veps^2}$,
is defined as in \req{eq:correlations} except that the denominator is to be supplemented by the
contribution from longitudinally polarized photons $\veps(|\Phi_{0+,0+}|^2 + |\Phi_{0-,0+}|^2)$. 
The $Q^2\to 0$ limit of $A_{LL}^{\cos{(0\varphi)}}/\sqrt{1-\veps^2}$ is
the $A_{LL}$ parameter we discussed in the previous subsection. The second modulation of $A_{LL}$ is related to the
amplitudes by
\be
\frac{A_{LL}^{P,\cos{\varphi}}}{\sqrt{\veps(1-\veps)}}\,\sigma_0^P \= - {\rm Re}\,\Big[ \big(\Phi^P_{0+,++}
  + \Phi^P_{0+,-+}\big)\,\Phi^{P*}_{0+,0+}
  + \big(\Phi^P_{0-,++} + \Phi^P_{0-,-+}\big)\,\Phi^{P*}_{0-,0+}\Big]\,,
\ee
where
\be
 \sigma_0^P \=  {\sum_{\nu',\mu}|\Phi^P_{0\nu',\mu +}|^2 + \veps \sum_{\nu'} |\Phi^P_{0\nu',0+}|^2}
 \ee
 is the unseparated cross section without the phase space factor.
 We stress that for both the modulations of $A_{LL}$ the longitudinal target polarization is defined relative to the
 direction of the virtual photon. In experiments the target polarization is usually defined relative to lepton beam
 direction. The transform from one definition to the other one is investigated in great detail in the work by
 Diehl and Sapeta \ci{diehl-sapeta}. According to that work, the following relation 
\be
A_{LL}^{P,l}\=\cos{\theta_\gamma} A_{LL}^{P} - \sin{\theta_\gamma}\,A_{LT}^{P}(\phi_S=0)
\label{eq:transform}
\ee
holds for both the modulations. The angle $\phi_S$ denotes the orientation of the transversal target spin vector
with respect to the lepton plane. The label $l$ stands for the target polarization defined with respect to the
lepton beam. The angle $\theta_\gamma$ describes the rotation in the lepton plane from the direction of the
incoming lepton  to that of the virtual photon. It is given by
\be
\cos{\theta_\gamma}\=\frac{1+y\gamma^2/2}{\sqrt{1+\gamma^2}}
\ee
where $\gamma=2 x_B m/Q$ and    $y=(s+Q^2-m^2)/(2 m E_{Lab}+m^2)$ ($x_B$ is Bjorken-x and $E_{Lab}$ the beam energy
in the Lab frame). Since this rotation requires information on the actual experiment which is not at our disposal,
we refrain from quoting $A^{P,l}_{LL}$. The correlations $A_{LL}^{P,\cos{(0\varphi)}}/\sqrt{1-\veps^2}$ and
$A_{LL}^{P,\cos{\varphi}}/\sqrt{\veps(1-\veps)}$ still depend on $\veps$ through $\sigma_0^P$. In order to make
predictions we have tentatively chosen $\veps=0.6$.  The two modulations of $A_{LL}$ are shown in Fig.\
\ref{fig:spin-pi-electro} for $\pi^+$ electroproduction at $Q^2=1$ and $2\,\gev^2$ and $s=10.3\,\gev^2$. In order
to facilitate the use of the possible transform \req{eq:transform} from the direction of the virtual photon to
that of the lepton beam we also show in Fig.\ \ref{fig:spin-pi-electro} the correlations between the lepton
helicity and transversal target polarization defined by
\ba
\frac{A_{LT}^{P,\cos{(0\varphi)}}(\phi_s=0)}{\sqrt{\veps(1-\veps)}}&=& - 
             \frac{ {\rm Re}\,\Big[\Phi^{P*}_{0+,++}\Phi^P_{0-,0+} - \Phi^{P*}_{0-,++} \Phi^P_{0+,0+}\Big]}{\sigma_0^P}\,, \nn\\
\frac{A_{LT}^{P,\cos{\varphi}}(\phi_s=0)}{\sqrt{1-\veps^2}}&=& - 
             \frac{ {\rm Re}\,\Big[\Phi^{P*}_{0+,++}\Phi^P_{0-,-+} - \Phi^{P*}_{0-,++} \Phi^P_{0+,-+}\Big]}{\sigma_0^P}\,,
\ea
which also appear in \req{eq:transform}.

A large variety of other spin-dependent observables exist for electroproduction of pions. Their study goes beyond the
scope of the present work. 
Notice that the single-spin asymmetries frequently require phase differences. They are all zero for our  LO study
of the handbag mechanism, the LO amplitudes \req{eq:amplitudes} and \req{eq:long-amplitudes} are real.

\clearpage
%%%%%%%%%%%%%%%%%%%%%%%%%%%%%%%%%%%%%%%%%%%%%%%%%%%%%%%%%%%%%%%%%%%%%%%%%%%%%%%%%%%%%%%%%%%%%
\section{Remarks on uncertainties}
\label{sec:error}
%%%%%%%%%%%%%%%%%%%%%%%%%%%%%%%%%%%%%%%%%%%%%%%%%%%%%%%%%%%%%%%%%%%%%%%%%%%%%%%%%%%%%%%%%%%%
There are two types of uncertainties. On the one hand, there are those resulting from the use of kinematics for
which the prerequisite of the handbag approach namely that the Mandelstam variables should be much larger than a
typical hadronic scale of order $1\,\gev^2$ is not sufficiently well respected. The analysis of the
available photoproduction data as well as the photo- and electroproduction data to be expected
in the near future force us to work in a kinematical situation in which the requirement of large Mandelstam
variables is only marginally respected. On the other hand, there are the typical parametric uncertainties
due to error-burdened parameters of the \da s and the form factors.

For Mandelstam variables which are not much larger than $\Lambda^2$ the problem arises how to match $s, t$ and $u$
for the hadronic process with the ones for the partonic subprocess, $\sh, \th$ and $\uh$. There are several
possibilities to match the kinematics. They lead to different numerical results which may be regarded as
different nucleon-mass corrections \ci{huang02} of order $m^2/s$.
  Thus, one possibility is to identify the two sets of variables as in \req{eq:sub-full}. Alternatively, one may use
  \be
  \th\=t\,, \qquad \sh\=s-m^2\,, \qquad \uh\=u-m^2\,,
  \ee
  which choice guarantees the mass-shell condition for the subprocess, $\sh+\uh+\th\=0$. There are other
  possibilities \ci{huang02}. For the kinematics of interest in this work the mass corrections due to the
  different matching recipes are large, in particular
  for $|\cos{\theta}|\to 1$. They amount to about $50\%$ for photo- and electroproduction. However, for ratios
  of cross sections which we mainly present for electroproduction these mass corrections are smaller. Thus,
  for $d\sigma_{T}(Q^2)/d\sigma_T(0)$ and $d\sigma_{TT}(Q^2)/d\sigma_T(Q^2)$ they only amount to about $20\%$ for not
  to large values of $Q^2$. For the helicity correlations, $A_{LL}$ and $K_{LL}$, which likewise represent ratios of
  cross sections, the mass corrections are less than $20\%$. 
  
  The parametric uncertainty of the twist-3 contribution is very large. As mentioned in Sec.\
  \ref{sec:DA} the parameters of the 3-body twist-3 \da{} have errors of about $30\%$. Together with the
  uncertainties of the large-$-t$ behavior of the transversity form factors the parametric uncertainties
  of the cross sections amount to about $70\%$ in the regions of twist-3 dominance. Evidently, for ratios
  of cross sections and for the helicity correlations the parametric uncertainties are much smaller since
  most of the uncertainties of the twist-3 \da{} cancel. 
The parametric uncertainty of the twist-2 contribution is insignificant compared to the other sources of errors. 
%%%%%%%%%%%%%%%%%%%%%%%%%%%%%%%%%%%%%%%%%%%%%%%%%%%%%%%%%%%%%%%%%%
\section{Summary}
\label{sec:summary}
%%%%%%%%%%%%%%%%%%%%%%%%%%%%%%%%%%%%%%%%%%%%%%%%%%%%%%%%%%%%%%%%%%
We have calculated wide-angle photo- and electroproduction of pions within the handbag factorization
scheme to twist-3 accuracy and LO of perturbative QCD. In this mechanism the amplitudes factorize into
hard partonic subprocess amplitudes and soft form factors representing $1/x$-moments of zero-skewness GPDs.
The twist-3 contributions to the subprocess amplitudes include the 2-body, $q\bar{q}$, as well as the
3-body, $q\bar{q}g$, Fock components of the pion. In light-cone gauge we are using for the vacuum-meson matrix
elements, the equation of motion which is formally an inhomogeneous linear first-order differential equation,
fixes the 2-body twist-3 \da s, $\phi_{\pi p}$ and $\phi_{\pi \sigma}$, for a given 3-body \da . Thus, only
two independent pion \da s remain as soft physics input to the handbag mechanism in addition to the form
factors, the usual ones, being related to the helicity non-flip GPDs, as well as the transversity form factors.
Taking the \da s and the form factors
from our previous work \ci{KPK18}, we evaluated the photoproduction cross sections within the handbag
mechanism and compared the results with experimental data. It seems that the energy dependence of the
handbag contribution is somewhat too strong. Including evolution and the $t$-dependence of the form factors
it is approximately $s^{-9}$ at fixed $\cos{\theta}$. Therefore, as an alternative we also presented results
evaluated at the fixed scale of $1\,\gev$ which agree fairly well with experiment. A better determination of
the large $-t$ behavior of the form factors is required before a final answer can be given what the energy
dependence of handbag predictions is. We also give detailed predictions for pion electroproduction and discuss
spin effects, especially helicity correlations. The gross features of wide-angle electroproduction bear
similarities to DVMP: the $\pi^0$ channel is dominated by the twist-3 contributions, for the $\pi^\pm$
channels twist-2 contributions matter in the forward hemisphere. Data on pion electroproduction would
allow to extract detailed information on the large $-t$-behavior of the form factors and the transversity
GPDs at zero skewness. This knowledge bears implications of our understanding of the parton densities in the
transverse position plane \ci{burkardt,diehl-haegler}. It may also be of help in understanding some of the
spin density matrix elements of deeply virtual vector-meson production \ci{GK4,GK7}.
In the present work, we restricted ourselves to the case of the pion. However, the generalization to other
pseudoscalar mesons is straightforward.

{\it Acknowledgements} 
We would like to thank Tania Robens and Lech Szymanowski
for discussions and comments.
This publication is supported 
by the Croatian Science Foundation project IP-2019-04-9709,
by the EU Horizon 2020 research and innovation programme, STRONG-2020
project, under grant agreement No 824093
and by Deutsche
Forschungsgemeinschaft (DFG) through the Research
Unit FOR 2926, “Next Generation pQCD for Hadron
Structure: Preparing for the EIC”, project number
40824754.

%%%%%%%%%%%%%%%%%%%%%%%%%%%%%%%%%%%%%%%%%%%%%%%%%%%%%%%%%%%%%%%%%
\begin{appendix}
\renewcommand{\theequation}{\Alph{section}.\arabic{equation}}%
\setcounter{equation}{0}
\section{The evolution of the pion \da s}
\label{sec:evolution}
In this appendix we compile the anomalous dimensions needed to evolve the \da s used in the previous sections.
The anomalous dimensions can be found in \ci{ball98,braun90}. The evolution goes with the quantity
\be
L\=\frac{\alpha_S(\mu_R)}{\alpha_S(\mu_O)}
  =\frac{\ln{(\muO/\Lambda^2_{\rm QCD})}}{\ln{(\muR/\Lambda^2_{\rm QCD})}} 
\label{eq:L}
\ee
The second Gegenbauer coefficient of the twist-2 pion \da{} evolves as
\be
a_2(\mu_R)\=a_2(\mu_0)\,L^{\gamma_2/\beta_0}
\ee
with $\gamma_2=50/9$ and $\beta_0=(11 N_C-2n_f)/3$. The mass parameter, $\mu_\pi$
evolves as
\be
\mu_\pi(\mu_R) \= L^{-4/\beta_0} \mu_\pi(\mu_0)\,.
\ee
 The parameters of the 3-body \da  evolve as
\ba
f_{3\pi}(\mu_F)&=&L^{(16/3 C_F-1)/\beta_0}\, f_{3\pi}(\mu_0)\,,  \nn\\
 \omega_{1,0}(\mu_R)&=& L^{(-25/6C_F+11/3C_A)/\beta_0}\omega_{1,0}(\mu_0)\,,  \nn\\
\omega_{11}(\mu_R) &=& \frac1{\gamma_+-\gamma_-}\,
            \left[(\gamma_--\gamma_{11}) A_+(\mu_0) L^{(\gamma_+-16/3 C_F+1)/\beta_0} \right. \nn\\
            &&\left. \hspace*{0.12\tw}  
            + (\gamma_+-\gamma_{11}) A_-(\mu_0) L^{(\gamma_--16/3 C_F+1)/\beta_0}\right]\,,  \nn\\ 
\omega_{20}(\mu_R) &=& \frac14\frac{\gamma_{21}}{\gamma_--\gamma_+}\,
    \left[A_+(\mu_0) L^{(\gamma_+-16/3C_F+1)/\beta_0}  \right. \nn\\
           && \left. \hspace*{0.13\tw} + A_-(\mu_0) L^{(\gamma_--16/3C_F+1)/\beta_0}\right]\,,
\ea
where  
\ba
A_+(\mu_0)&=& -\omega_{11}(\mu_0) - 4\frac{\gamma_+-\gamma_{11}}{\gamma_{21}}\omega_{20}(\mu_0)\,, \nn\\
A_-(\mu_0)&=& \phantom{-}\omega_{11}(\mu_0) + 4\frac{\gamma_--\gamma_{11}}
                                             {\gamma_{21}}\omega_{20}(\mu_0)\,.   
\ea
The anomalous dimensions are
\be
\gamma_{11}\=\frac{122}{9}\,, \quad \gamma_{22}\=\frac{511}{45}\,, \quad \gamma_{12}\=\frac53\,,
\quad \gamma_{21}\=\frac{21}{5}\,,
\ee
with the eigenvalues
\be
\gamma_\pm \= \frac12\big[\gamma_{11}+\gamma_{22}
               \pm\sqrt{(\gamma_{11}-\gamma_{22})^2+4\gamma_{12}\gamma_{21}}\;\Big]\,.
\ee
\end{appendix}
%%%%%%%%%%%%%%%%%%%%%%%%%%%%%%%%%%%%%%%%%%%%%%%%%%%%%%%%%%%%%%%%%
%%%%%%%%%%%%%%%%%%%%%%%%%%%%%%%%%%%%%%%%%%%%%%%%%%%%%%%%%%%%%%%%%%%%%

\end{document}